\documentclass{llncs}

\pdfoutput=1

\usepackage{isabelle}
\usepackage{isabellesym}
\usepackage[hyphens]{url}
\usepackage{pdfsetup}

\usepackage{microtype}

\usepackage{latexsym}
\usepackage{amssymb}
\usepackage{stmaryrd}

\renewcommand{\isasymlbrakk}{\isamath{\llbracket\,}}
\renewcommand{\isasymrbrakk}{\isamath{\,\rrbracket}}
\renewcommand{\isasymlbrace}{\isamath{\mathopen{\lbrace\mkern-4mu\mid}}}
\renewcommand{\isasymrbrace}{\isamath{\mathclose{\mid\mkern-4mu\rbrace}}}
\renewcommand{\isasymlparr}{\isamath{\llparenthesis\mkern1mu}}
\renewcommand{\isasymrparr}{\isamath{\mkern1mu\rrparenthesis}}

\usepackage{amsmath}

\renewenvironment{isamarkuptext}{\par}{}

\renewenvironment{isabellebody}{}{}

\renewenvironment{isabelle}
    {\begin{center}\isaspacing\isastyle}
    {\end{center}}

\isabellestyle{it}

\begin{document}

\title{A Process Calculus\\for Formally Verifying\\Blockchain Consensus Protocols}
\titlerunning{A Process Calculus for Blockchain Consensus Protocols}
\author{Wolfgang Jeltsch\inst{1,2}\orcidID{0000-0002-8068-8401}}
\institute{Well-Typed\and IOHK\\\email{wolfgang@well-typed.com}}
\maketitle

\begin{abstract}

Blockchains are becoming increasingly relevant in a variety of fields, such as finance, logistics,
and real estate. The fundamental task of a blockchain system is to establish data consistency among
distributed agents in an open network. Blockchain consensus protocols are central for performing
this task.

Since consensus protocols play such a crucial role in blockchain technology, several projects are
underway that apply formal methods to these protocols. One such project is carried out by a team of
the Formal Methods Group at IOHK. This project, in which the author is involved, aims at a formally
verified implementation of the Ouroboros family of consensus protocols, the backbone of the Cardano
blockchain. The first outcome of our project is the $\natural$-calculus (pronounced ``natural
calculus''), a general-purpose process calculus that serves as our implementation language. The
$\natural$-calculus is a domain-specific language embedded in a functional host language using
higher-order abstract syntax.

This paper will be a ramble through the $\natural$-calculus. First we will look at its language and
its operational semantics. The latter is unique in that it uses a stack of two labeled transition
systems to treat phenomena like data transfer and the opening and closing of channel scope in a
modular fashion. The presence of multiple transition systems calls for a generic treatment of
derived concurrency concepts. We will see how such a treatment can be achieved by capturing notions
like scope opening and silent transitions abstractly using axiomatically defined algebraic
structures based on functors and monads.


\end{abstract}

\begin{isabellebody}%
\setisabellecontext{Content}%
\isadeliminvisible
\endisadeliminvisible
\isataginvisible
\isacommand{theory}\isamarkupfalse%
\ Content\isanewline
\ \ \isakeyword{imports}\ Chi{\isacharunderscore}Calculus{\isachardot}Proper{\isacharunderscore}Weak{\isacharunderscore}Transition{\isacharunderscore}System\isanewline
\isakeyword{begin}%
\endisataginvisible
{\isafoldinvisible}%
\isadeliminvisible
\endisadeliminvisible
\isadelimdocument
\endisadelimdocument
\isatagdocument
\isamarkupsection{Introduction%
}
\isamarkuptrue%
\endisatagdocument
{\isafolddocument}%
\isadelimdocument
\endisadelimdocument
\begin{isamarkuptext}%
A blockchain is an open, distributed database that stores a growing list, the \emph{ledger}, and
achieves security by employing advanced cryptographic methods. Blockchains are used in finance for
implementing cryptocurrencies and smart contracts and have applications in other fields too.

A blockchain system establishes data consistency using a \emph{consensus protocol}. There are two main
kinds of such protocols:

\begin{itemize}%
\item \emph{Proof-of-work} protocols require participants to solve computational puzzles in order to
contribute data to the blockchain.

\item \emph{Proof-of-stake} protocols make the opportunity to contribute data dependent on the stake
participants possess, such as money in a cryptocurrency.%
\end{itemize}

Since the correctness of a blockchain system rests on the correctness of its consensus protocol,
several projects are underway that apply formal methods to consensus protocols. One such project
is carried out by a team of the Formal Methods Group at IOHK. This project, in which the author is
involved, aims at a formally verified implementation of the Ouroboros family of consensus
protocols~\cite{kiayias:crypto-2017,david:eurocrypt-2018,badertscher:ccs-2018},
which form the backbone of the Cardano blockchain.

All protocols in the Ouroboros family use the proof-of-stake mechanism and come with rigorous
security guarantees. In fact, the original Ouroboros protocol, dubbed Ouroboros Classic, was the
first proof-of-stake protocol to have such guarantees. The Cardano blockchain is the basis of the
cryptocurrency Ada and the smart contract languages Plutus~\cite{chakravarty:plutus} and
Marlowe~\cite{lamela:isola-2018}. Both Plutus and Marlowe are functional languages, but while
Plutus is Turing-complete, Marlowe is deliberately restricted in its expressivity to make
implementing common contracts easy.

In this paper, we report on the first outcome of our Ouroboros formalization effort: the
\emph{\isa{{\isasymnatural}}-calculus} (pronounced ``natural calculus''). The \isa{{\isasymnatural}}-calculus is a process calculus that
serves as our specification and implementation language. We make the following contributions:

\begin{itemize}%
\item In Sect.~\ref{the-natural-calculus}, we present the language and the operational semantics of
the \isa{{\isasymnatural}}-calculus. The latter is unique in that it uses a stack of two labeled transition
systems to treat phenomena like data transfer and the opening and closing of channel scope in
a modular fashion

\item The presence of multiple transition systems calls for a generic treatment of derived
concurrency concepts, such as strong and weak bisimilarity. In
Sect.~\ref{residuals-axiomatically}, we develop an abstract theory of transition systems to
achieve such a generic treatment. Our theory captures notions like scope opening and silent
transitions using axiomatically defined algebraic structures. In these structures, functors
and monads play a crucial role.%
\end{itemize}

\noindent We conclude this paper with Sects. \ref{related-work} and~\ref{summary-and-outlook}, where we
discuss related work and give a summary and an outlook.

To this end, we have formalized~\cite{jeltsch:ouroboros-formalization} large parts of the
\isa{{\isasymnatural}}-calculus and our complete theory of transition systems in Isabelle/HOL. Furthermore, we have
produced this paper from documented Isabelle source code~\cite{jeltsch:wflp-2019-source}, which
we have checked against our formalization.%
\end{isamarkuptext}\isamarkuptrue%
\isadelimdocument
\endisadelimdocument
\isatagdocument
\isamarkupsection{The \isa{{\isasymnatural}}-Calculus%
}
\isamarkuptrue%
\endisatagdocument
{\isafolddocument}%
\isadelimdocument
\endisadelimdocument
\label{the-natural-calculus}
\begin{isamarkuptext}%
The \isa{{\isasymnatural}}-calculus is a process calculus in the tradition of the
  \isa{{\isasympi}}-calculus~\cite{milner:pi-calculus}. It is not tied to blockchains in any way but is a
  universal language for concurrent and distributed computing.

  Unlike the \isa{{\isasympi}}-calculus, the \isa{{\isasymnatural}}-calculus is not an isolated language but is embedded into
  functional host languages. In our application scenario, we use embeddings into both Haskell, for
  execution, and Isabelle/HOL, for verification. The user is expected to write programs as
  Haskell-embedded process calculus terms, which can then be turned automatically into
  Isabelle-embedded process calculus terms to make them available for verification. In this paper,
  we focus on the Isabelle embedding, leaving the discussion of the Haskell embedding for another
  time. Whenever we use the term ``\isa{{\isasymnatural}}-calculus'', we refer to either the calculus in general or its
  embedding into Isabelle/HOL.

  Our embedding technique uses higher-order abstract syntax (HOAS)~\cite{pfenning:pldi-1988},
  which means we represent binding of names using functions of the host language. An immediate
  consequence of this is that the host language deals with all the issues regarding names, like
  shadowing and \isa{{\isasymalpha}}-equivalence, which simplifies the implementation of the calculus. Furthermore,
  HOAS gives us support for arbitrary data for free, since we can easily represent data by values of
  the host language. This lifts the restriction of the \isa{{\isasympi}}-calculus that channels are the only kind
  of data. Finally, HOAS allows us to move computation, branching, and recursion to the host
  language level and thus further simplify the implementation of the calculus.

  The \isa{{\isasymnatural}}-calculus is similar to \isa{{\isasympsi}}-calculi~\cite{bengtson:lmcs-7-1} in that it adds support for
  arbitrary data to the core features of the \isa{{\isasympi}}-calculus. However, since the \isa{{\isasymnatural}}-calculus uses
  HOAS, we can avoid much of the complexity of \isa{{\isasympsi}}-calculi that comes from their need to cope with
  data-related issues themselves.%
\end{isamarkuptext}\isamarkuptrue%
\isadelimdocument
\endisadelimdocument
\isatagdocument
\isamarkupsubsection{Language%
}
\isamarkuptrue%
\endisatagdocument
{\isafolddocument}%
\isadelimdocument
\endisadelimdocument
\begin{isamarkuptext}%
We define a coinductive data type \isa{process} whose values are the terms of the \isa{{\isasymnatural}}-calculus.
We call these terms simply \emph{processes}.

In the following, we list the different kinds of processes. For describing their syntax, we use
statements of the form \isa{C\ x\isactrlsub {\isadigit{1}}\ {\isasymdots}\ x\isactrlsub n\ {\isasymequiv}\ e}. The left-hand side of such a statement is an application
of a data constructor of the \isa{process} type to argument variables; it showcases the ordinary
notation for the respective kind of processes. The right-hand side is a term that is equal to the
left-hand side but uses convenient notation introduced by us using Isabelle's means for defining
custom syntax. The kinds of processes are as follows:

\begin{itemize}%
\item Do nothing:%
\begin{isabelle}%
{\isachardoublequote}Stop\ {\isasymequiv}\ {\isasymzero}{\isachardoublequote}%
\end{isabelle}

\item Send value~\isa{x} to channel~\isa{a}:%
\begin{isabelle}%
{\isachardoublequote}Send\ a\ x\ {\isasymequiv}\ a\ {\isasymtriangleleft}\ x{\isachardoublequote}%
\end{isabelle}

\item Receive value~\isa{x} from channel~\isa{a} and continue with \isa{P\ x}:%
\begin{isabelle}%
{\isachardoublequote}Receive\ a\ P\ {\isasymequiv}\ a\ {\isasymtriangleright}\ x{\isachardot}\ P\ x{\isachardoublequote}%
\end{isabelle}

\item Perform processes \isa{p} and~\isa{q} concurrently:%
\begin{isabelle}%
{\isachardoublequote}Parallel\ p\ q\ {\isasymequiv}\ p\ {\isasymparallel}\ q{\isachardoublequote}%
\end{isabelle}

\item Create a new channel~\isa{a} and continue with \isa{P\ a}:%
\begin{isabelle}%
{\isachardoublequote}NewChannel\ P\ {\isasymequiv}\ {\isasymnu}\ a{\isachardot}\ P\ a{\isachardoublequote}%
\end{isabelle}%
\end{itemize}

\noindent The binders (\isa{{\isasymtriangleright}}~and~\isa{{\isasymnu}}) bind stronger than the infix operator~(\isa{{\isasymparallel}}), which is not what the
reader might have expected but is typical for process calculi.

There are a few interesting points to note regarding processes and their notation:

\begin{itemize}%
\item Our use of HOAS manifests itself in the \isa{Receive} and \isa{NewChannel} cases. In both
of them, the respective data constructor takes an argument \isa{P} that is a continuation
which maps a received value or a newly created channel to a remainder process.

\item Although dependencies on received values and newly created channels are encoded using
functions, we can still use convenient binder notation for \isa{Receive} and
\isa{NewChannel} processes. A term~\isa{e} in \isa{a\ {\isasymtriangleright}\ x{\isachardot}\ e} or \isa{{\isasymnu}\ a{\isachardot}\ e}
does not have to be an application of a function~\isa{P} to the bound variable. Every term
that possibly mentions the bound variable is fine. For example, \isa{a\ {\isasymtriangleright}\ x{\isachardot}\ b\ {\isasymtriangleleft}\ x\ {\isasymparallel}\ c\ {\isasymtriangleleft}\ x} is
a valid term, which is equal to \isa{{\isachardoublequote}Receive\ a\ {\isacharparenleft}{\isasymlambda}x{\isachardot}\ b\ {\isasymtriangleleft}\ x\ {\isasymparallel}\ c\ {\isasymtriangleleft}\ x{\isacharparenright}{\isachardoublequote}}.

\item HOAS gives us the opportunity to construct processes that include computation and branching,
despite the process calculus not having dedicated constructs for these things. For example,
the process \isa{{\isachardoublequote}a\ {\isasymtriangleright}\ y{\isachardot}\ {\isacharparenleft}if\ y\ {\isasymnoteq}\ x\ then\ b\ {\isasymtriangleleft}\ y\ else\ {\isasymzero}{\isacharparenright}{\isachardoublequote}}, which performs a kind of
conditional forwarding, carries the inequality test and the branching inside the continuation
argument of \isa{Receive}.

\item \isa{{\isachardoublequote}Send{\isachardoublequote}} does not have a continuation argument. This is to make communication
effectively asynchronous. The operational semantics defines communication in the usual way,
making it actually synchronous, but without \isa{{\isachardoublequote}Send{\isachardoublequote}} continuations, synchrony
cannot be observed. This approach is common for asynchronous process calculi and is used, for
example, in the asynchronous \isa{{\isasympi}}-calculus~\cite{boudol:inria-00076939}. We use asynchronous
communication, because it is sufficient for our use case and easier to implement in common
programming languages, like Haskell.

\item The \isa{{\isasymnatural}}-calculus does not have a construct for nondeterministic choice, because execution of
nondeterministic choice is difficult to implement.

\item The \isa{{\isasymnatural}}-calculus does not have a construct for replication. We do not need such a construct,
since the \isa{process} type is coinductive and thus allows us to form infinite terms. The
replication of a process~\isa{p} can be defined as the infinite term \isa{p\ {\isasymparallel}\ p\ {\isasymparallel}\ {\isasymdots}}, that is, the
single term \isa{p\isactrlsup {\isasyminfinity}} for which \isa{p\isactrlsup {\isasyminfinity}\ {\isacharequal}\ p\ {\isasymparallel}\ p\isactrlsup {\isasyminfinity}}.%
\end{itemize}%
\end{isamarkuptext}\isamarkuptrue%
\isadelimdocument
\endisadelimdocument
\isatagdocument
\isamarkupsubsection{Operational Semantics%
}
\isamarkuptrue%
\endisatagdocument
{\isafolddocument}%
\isadelimdocument
\endisadelimdocument
\label{operational-semantics}
\isadeliminvisible
\endisadeliminvisible
\isataginvisible
\isacommand{no{\isacharunderscore}notation}\isamarkupfalse%
\ proper{\isacharunderscore}transition\ {\isacharparenleft}\isakeyword{infix}\ {\isachardoublequoteopen}{\isasymrightarrow}\isactrlsub {\isasymsharp}{\isachardoublequoteclose}\ {\isadigit{5}}{\isadigit{0}}{\isacharparenright}\isanewline
\isacommand{notation}\isamarkupfalse%
\ proper{\isacharunderscore}transition\ {\isacharparenleft}{\isachardoublequoteopen}{\isacharunderscore}\ {\isasymrightarrow}{\isacharunderscore}{\isachardoublequoteclose}\ {\isacharbrackleft}{\isadigit{5}}{\isadigit{1}}{\isacharcomma}\ {\isadigit{5}}{\isadigit{1}}{\isacharbrackright}\ {\isadigit{5}}{\isadigit{0}}{\isacharparenright}%
\endisataginvisible
{\isafoldinvisible}%
\isadeliminvisible
\endisadeliminvisible
\begin{isamarkuptext}%
We define the operational semantics of the \isa{{\isasymnatural}}-calculus as a labeled transition system. We write
\isa{p\ {\isasymrightarrow}{\isasymlparr}{\isasymxi}{\isasymrparr}\ q} to say that \isa{p} can transition to~\isa{q} with label~\isa{{\isasymxi}}.

We handle isolated sending and receiving as well as communication in the standard manner. We
introduce labels \isa{a\ {\isasymtriangleleft}\ x}, \isa{a\ {\isasymtriangleright}\ x}, and
\isa{{\isasymtau}}, which denote sending of a value~\isa{x} to a channel~\isa{a},
receiving of a value~\isa{x} from a channel~\isa{a}, and internal communication, respectively,
and call these labels \emph{actions}. Then we introduce the following rules:

\begin{itemize}%
\item Sending:%
\begin{isabelle}%
a\ {\isasymtriangleleft}\ x\ {\isasymrightarrow}{\isasymlparr}a\ {\isasymtriangleleft}\ x{\isasymrparr}\ {\isasymzero}%
\end{isabelle}

\item Receiving:%
\begin{isabelle}%
a\ {\isasymtriangleright}\ x{\isachardot}\ P\ x\ {\isasymrightarrow}{\isasymlparr}a\ {\isasymtriangleright}\ x{\isasymrparr}\ P\ x%
\end{isabelle}

\item Communication:%
\begin{isabelle}%
{\isasymlbrakk}p\ {\isasymrightarrow}{\isasymlparr}a\ {\isasymtriangleleft}\ x{\isasymrparr}\ p{\isacharprime}{\isacharsemicolon}\ q\ {\isasymrightarrow}{\isasymlparr}a\ {\isasymtriangleright}\ x{\isasymrparr}\ q{\isacharprime}{\isasymrbrakk}\ {\isasymLongrightarrow}\ p\ {\isasymparallel}\ q\ {\isasymrightarrow}{\isasymlparr}{\isasymtau}{\isasymrparr}\ p{\isacharprime}\ {\isasymparallel}\ q{\isacharprime}%
\end{isabelle}

\item Acting within a subsystem:%
\begin{isabelle}%
p\ {\isasymrightarrow}{\isasymlparr}{\isasymxi}{\isasymrparr}\ p{\isacharprime}\ {\isasymLongrightarrow}\ p\ {\isasymparallel}\ q\ {\isasymrightarrow}{\isasymlparr}{\isasymxi}{\isasymrparr}\ p{\isacharprime}\ {\isasymparallel}\ q%
\end{isabelle}%
\end{itemize}

\noindent The last two of these rules have symmetric versions, which we do not show here for the sake of
simplicity.%
\end{isamarkuptext}\isamarkuptrue%
\isadeliminvisible
\endisadeliminvisible
\isataginvisible
\isacommand{axiomatization}\isamarkupfalse%
\ chan{\isacharunderscore}to{\isacharunderscore}val\ {\isacharcolon}{\isacharcolon}\ {\isachardoublequoteopen}chan\ {\isasymRightarrow}\ val{\isachardoublequoteclose}\ {\isacharparenleft}{\isachardoublequoteopen}{\isasymcent}{\isacharunderscore}{\isachardoublequoteclose}{\isacharparenright}%
\endisataginvisible
{\isafoldinvisible}%
\isadeliminvisible
\endisadeliminvisible
\renewcommand{\isasymcent}{}
\isadeliminvisible
\endisadeliminvisible
\isataginvisible
\isacommand{lemma}\isamarkupfalse%
\ pi{\isacharunderscore}calculus{\isacharunderscore}closing{\isacharcolon}\isanewline
\ \ \isakeyword{assumes}\ {\isachardoublequoteopen}p\ {\isasymrightarrow}{\isasymlparr}a\ {\isasymtriangleleft}\ {\isasymnu}\ b{\isachardot}\ {\isasymcent}b{\isasymrparr}\ P\ b{\isachardoublequoteclose}\ \isakeyword{and}\ {\isachardoublequoteopen}{\isasymAnd}b{\isachardot}\ q\ {\isasymrightarrow}{\isasymlparr}a\ {\isasymtriangleright}\ {\isasymcent}b{\isasymrparr}\ Q\ b{\isachardoublequoteclose}\isanewline
\ \ \isakeyword{shows}\ {\isachardoublequoteopen}p\ {\isasymparallel}\ q\ {\isasymrightarrow}{\isasymlparr}{\isasymtau}{\isasymrparr}\ {\isasymnu}\ b{\isachardot}\ {\isacharparenleft}P\ b\ {\isasymparallel}\ Q\ b{\isacharparenright}{\isachardoublequoteclose}\isanewline
\isacommand{proof}\isamarkupfalse%
\ {\isacharminus}\isanewline
\ \ \isacommand{from}\isamarkupfalse%
\ assms{\isacharparenleft}{\isadigit{1}}{\isacharparenright}\ \isacommand{obtain}\isamarkupfalse%
\ U\ \isakeyword{where}\ {\isachardoublequoteopen}p\ {\isasymrightarrow}\isactrlsub {\isasymflat}{\isasymlbrace}{\isasymnu}\ b{\isasymrbrace}\ U\ b{\isachardoublequoteclose}\ \isakeyword{and}\ {\isachardoublequoteopen}{\isasymAnd}b{\isachardot}\ U\ b\ {\isasymrightarrow}\isactrlsub {\isasymflat}{\isasymlbrace}a\ {\isasymtriangleleft}\ {\isasymcent}b{\isasymrbrace}\ P\ b{\isachardoublequoteclose}\isanewline
\ \ \ \ \isacommand{by}\isamarkupfalse%
\ {\isacharparenleft}fast\ elim{\isacharcolon}\ proper{\isacharunderscore}transition{\isachardot}cases{\isacharparenright}\isanewline
\ \ \isacommand{moreover}\isamarkupfalse%
\ \isacommand{from}\isamarkupfalse%
\ assms{\isacharparenleft}{\isadigit{2}}{\isacharparenright}\ \isacommand{have}\isamarkupfalse%
\ {\isachardoublequoteopen}{\isasymAnd}b{\isachardot}\ q\ {\isasymrightarrow}\isactrlsub {\isasymflat}{\isasymlbrace}a\ {\isasymtriangleright}\ {\isasymcent}b{\isasymrbrace}\ Q\ b{\isachardoublequoteclose}\isanewline
\ \ \ \ \isacommand{by}\isamarkupfalse%
\ {\isacharparenleft}fastforce\ elim{\isacharcolon}\ proper{\isacharunderscore}transition{\isachardot}cases{\isacharparenright}\isanewline
\ \ \isacommand{ultimately}\isamarkupfalse%
\ \isacommand{show}\isamarkupfalse%
\ {\isacharquery}thesis\isanewline
\ \ \ \ \isacommand{by}\isamarkupfalse%
\ {\isacharparenleft}fastforce\ intro{\isacharcolon}\ ltr\ communication\ opening{\isacharunderscore}left\ scoped{\isacharunderscore}acting\ simple{\isacharparenright}\isanewline
\isacommand{qed}\isamarkupfalse%
\endisataginvisible
{\isafoldinvisible}%
\isadeliminvisible
\endisadeliminvisible
\begin{isamarkuptext}%
Channels created by \isa{NewChannel} are initially local. However, such channels can later be
made visible by sending them to other subsystems. Let us see how this is captured by the
transition system of the \isa{{\isasympi}}-calculus. Besides ordinary sending labels
\isa{a\ {\isasymtriangleleft}\ b}, the \isa{{\isasympi}}-calculus has labels \isa{a\ {\isasymtriangleleft}\ {\isasymnu}\ b{\isachardot}\ b} that additionally
bind the variable~\isa{b}. The bound variable denotes a channel not yet known to the outside.
Using it as the value being sent thus conveys the information that a local channel is being
published by sending it to~\isa{a}. When used as part of a transition statement, the scope of the
binder includes the target process, so that the target process can depend on the published
channel. Therefore, the general form of a transition statement with local channel publication is
\isa{p\ {\isasymrightarrow}{\isasymlparr}a\ {\isasymtriangleleft}\ {\isasymnu}\ b{\isachardot}\ {\isasymcent}b{\isasymrparr}\ Q\ b}. The following rules are HOAS versions of the \isa{{\isasympi}}-calculus rules that
deal with local channels:

\begin{itemize}%
\item Scope opening:%
\begin{isabelle}%
{\isacharparenleft}{\isasymAnd}b{\isachardot}\ P\ b\ {\isasymrightarrow}{\isasymlparr}a\ {\isasymtriangleleft}\ {\isasymcent}b{\isasymrparr}\ Q\ b{\isacharparenright}\ {\isasymLongrightarrow}\ {\isasymnu}\ b{\isachardot}\ P\ b\ {\isasymrightarrow}{\isasymlparr}a\ {\isasymtriangleleft}\ {\isasymnu}\ b{\isachardot}\ {\isasymcent}b{\isasymrparr}\ Q\ b%
\end{isabelle}

\item Communication with scope closing:%
\begin{isabelle}%
{\isasymlbrakk}p\ {\isasymrightarrow}{\isasymlparr}a\ {\isasymtriangleleft}\ {\isasymnu}\ b{\isachardot}\ {\isasymcent}b{\isasymrparr}\ P\ b{\isacharsemicolon}\ {\isasymAnd}b{\isachardot}\ q\ {\isasymrightarrow}{\isasymlparr}a\ {\isasymtriangleright}\ {\isasymcent}b{\isasymrparr}\ Q\ b{\isasymrbrakk}\ {\isasymLongrightarrow}\ p\ {\isasymparallel}\ q\ {\isasymrightarrow}{\isasymlparr}{\isasymtau}{\isasymrparr}\ {\isasymnu}\ b{\isachardot}\ {\isacharparenleft}P\ b\ {\isasymparallel}\ Q\ b{\isacharparenright}%
\end{isabelle}

\item Acting inside scope:%
\begin{isabelle}%
{\isacharparenleft}{\isasymAnd}a{\isachardot}\ P\ a\ {\isasymrightarrow}{\isasymlparr}{\isasymdelta}{\isasymrparr}\ Q\ a{\isacharparenright}\ {\isasymLongrightarrow}\ {\isasymnu}\ a{\isachardot}\ P\ a\ {\isasymrightarrow}{\isasymlparr}{\isasymdelta}{\isasymrparr}\ {\isasymnu}\ a{\isachardot}\ Q\ a%
\end{isabelle}%
\end{itemize}

For the \isa{{\isasymnatural}}-calculus, these rules are unfortunately not enough. Unlike the \isa{{\isasympi}}-calculus, the
\isa{{\isasymnatural}}-calculus permits arbitrary data to be sent, which includes values that contain several
channels, like pairs of channels and lists of channels. As a result, several local channels can be
published at once. Variants of the above rules that account for this possibility are complex and
hard to get right. The complexity has two reasons:

\begin{itemize}%
\item Some labels deal with multiple concepts, namely scope opening and sending. In the
\isa{{\isasymnatural}}-calculus, these labels are not necessarily of the relatively simple form \isa{a\ {\isasymtriangleleft}\ {\isasymnu}\ b{\isachardot}\ b}
discussed above, but generally of the more complex form \isa{{\isasymnu}\ b\isactrlsub {\isadigit{1}}\ {\isasymdots}\ b\isactrlsub n{\isachardot}\ a\ {\isasymtriangleleft}\ f\ b\isactrlsub {\isadigit{1}}\ {\isasymdots}\ b\isactrlsub n}, because
arbitrary values depending on multiple local channels can be sent.

\item Some rules deal with multiple concepts, namely the rule about communication with scope
closing, which deals with precisely these two things, and the rule about acting inside scope,
which essentially adds scope opening before and scope closing after the given action.%
\end{itemize}%
\end{isamarkuptext}\isamarkuptrue%
\isadeliminvisible
\endisadeliminvisible
\isataginvisible
\isacommand{no{\isacharunderscore}notation}\isamarkupfalse%
\ basic{\isacharunderscore}transition\ {\isacharparenleft}\isakeyword{infix}\ {\isachardoublequoteopen}{\isasymrightarrow}\isactrlsub {\isasymflat}{\isachardoublequoteclose}\ {\isadigit{5}}{\isadigit{0}}{\isacharparenright}\isanewline
\isacommand{notation}\isamarkupfalse%
\ basic{\isacharunderscore}transition\ {\isacharparenleft}{\isachardoublequoteopen}{\isacharunderscore}\ {\isasymrightarrow}\isactrlsub {\isasymflat}{\isacharunderscore}{\isachardoublequoteclose}\ {\isacharbrackleft}{\isadigit{5}}{\isadigit{1}}{\isacharcomma}\ {\isadigit{5}}{\isadigit{1}}{\isacharbrackright}\ {\isadigit{5}}{\isadigit{0}}{\isacharparenright}\isanewline
\isanewline
\isacommand{no{\isacharunderscore}notation}\isamarkupfalse%
\ proper{\isacharunderscore}transition\ {\isacharparenleft}{\isachardoublequoteopen}{\isacharunderscore}\ {\isasymrightarrow}{\isacharunderscore}{\isachardoublequoteclose}\ {\isacharbrackleft}{\isadigit{5}}{\isadigit{1}}{\isacharcomma}\ {\isadigit{5}}{\isadigit{1}}{\isacharbrackright}\ {\isadigit{5}}{\isadigit{0}}{\isacharparenright}\isanewline
\isacommand{notation}\isamarkupfalse%
\ proper{\isacharunderscore}transition\ {\isacharparenleft}{\isachardoublequoteopen}{\isacharunderscore}\ {\isasymrightarrow}\isactrlsub {\isasymsharp}{\isacharunderscore}{\isachardoublequoteclose}\ {\isacharbrackleft}{\isadigit{5}}{\isadigit{1}}{\isacharcomma}\ {\isadigit{5}}{\isadigit{1}}{\isacharbrackright}\ {\isadigit{5}}{\isadigit{0}}{\isacharparenright}%
\endisataginvisible
{\isafoldinvisible}%
\isadeliminvisible
\endisadeliminvisible
\begin{isamarkuptext}%
To tame this complexity, we conduct the definition of the transition system in two steps:

\begin{enumerate}%
\item We define a transition system that uses distinct transitions for opening scopes, so that each
label and each rule deals with a single concept only. We call this transition system the
\emph{basic transition system} and write a transition in this system \isa{p\ {\isasymrightarrow}\isactrlsub {\isasymflat}{\isasymlbrace}{\isasymxi}{\isasymrbrace}\ q}.

\item We define the transition system that describes the actual semantics of the \isa{{\isasymnatural}}-calculus by
adding a layer on top of the basic transition system that bundles scope opening and sending
transitions. We call this transition system the \emph{proper transition system} and write a
transition in this system \isa{p\ {\isasymrightarrow}\isactrlsub {\isasymsharp}{\isasymlparr}{\isasymxi}{\isasymrparr}\ q}.%
\end{enumerate}

The basic transition system has \emph{action labels} \isa{a\ {\isasymtriangleleft}\ x},
\isa{a\ {\isasymtriangleright}\ x}, and \isa{{\isasymtau}} as well as \emph{opening labels}~\isa{{\isasymnu}\ a},
the latter binding their variables in any following target process. The rules for sending,
receiving, and communication are the ones we have seen at the beginning of
\hyperref[operational-semantics]{this subsection}. For dealing with local channels, the basic
transition system contains the following rules:

\begin{itemize}%
\item Scope opening:%
\begin{isabelle}%
{\isasymnu}\ a{\isachardot}\ P\ a\ {\isasymrightarrow}\isactrlsub {\isasymflat}{\isasymlbrace}{\isasymnu}\ a{\isasymrbrace}\ P\ a%
\end{isabelle}

\item Scope closing after acting:%
\begin{isabelle}%
{\isasymlbrakk}p\ {\isasymrightarrow}\isactrlsub {\isasymflat}{\isasymlbrace}{\isasymnu}\ a{\isasymrbrace}\ Q\ a{\isacharsemicolon}\ {\isasymAnd}a{\isachardot}\ Q\ a\ {\isasymrightarrow}\isactrlsub {\isasymflat}{\isasymlbrace}{\isasymalpha}{\isasymrbrace}\ R\ a{\isasymrbrakk}\ {\isasymLongrightarrow}\ p\ {\isasymrightarrow}\isactrlsub {\isasymflat}{\isasymlbrace}{\isasymalpha}{\isasymrbrace}\ {\isasymnu}\ a{\isachardot}\ R\ a%
\end{isabelle}

\item Scope closing after another scope opening:%
\begin{isabelle}%
{\isasymlbrakk}p\ {\isasymrightarrow}\isactrlsub {\isasymflat}{\isasymlbrace}{\isasymnu}\ a{\isasymrbrace}\ Q\ a{\isacharsemicolon}\ {\isasymAnd}a{\isachardot}\ Q\ a\ {\isasymrightarrow}\isactrlsub {\isasymflat}{\isasymlbrace}{\isasymnu}\ b{\isasymrbrace}\ R\ a\ b{\isasymrbrakk}\ {\isasymLongrightarrow}\ p\ {\isasymrightarrow}\isactrlsub {\isasymflat}{\isasymlbrace}{\isasymnu}\ b{\isasymrbrace}\ {\isasymnu}\ a{\isachardot}\ R\ a\ b%
\end{isabelle}

\item Scope opening within a subsystem:%
\begin{isabelle}%
p\ {\isasymrightarrow}\isactrlsub {\isasymflat}{\isasymlbrace}{\isasymnu}\ a{\isasymrbrace}\ P\ a\ {\isasymLongrightarrow}\ p\ {\isasymparallel}\ q\ {\isasymrightarrow}\isactrlsub {\isasymflat}{\isasymlbrace}{\isasymnu}\ a{\isasymrbrace}\ P\ a\ {\isasymparallel}\ q%
\end{isabelle}%
\end{itemize}

\noindent The last rule has a symmetric version, which we do not show here for the sake of simplicity.

The proper transition system has labels \isa{a\ {\isasymtriangleright}\ x}, \isa{{\isasymtau}},
and \isa{a\ {\isasymtriangleleft}\ {\isasymnu}\ b\isactrlsub {\isadigit{1}}\ {\isasymdots}\ b\isactrlsub n{\isachardot}\ f\ b\isactrlsub {\isadigit{1}}\ {\isasymdots}\ b\isactrlsub n}, the latter binding their variables also in any following target
process. The rules for sending, receiving, and communication just refer to the basic transition
system:

\begin{itemize}%
\item Sending:%
\begin{isabelle}%
p\ {\isasymrightarrow}\isactrlsub {\isasymflat}{\isasymlbrace}a\ {\isasymtriangleleft}\ x{\isasymrbrace}\ q\ {\isasymLongrightarrow}\ p\ {\isasymrightarrow}\isactrlsub {\isasymsharp}{\isasymlparr}a\ {\isasymtriangleleft}\ x{\isasymrparr}\ q%
\end{isabelle}

\item Receiving:%
\begin{isabelle}%
p\ {\isasymrightarrow}\isactrlsub {\isasymflat}{\isasymlbrace}a\ {\isasymtriangleright}\ x{\isasymrbrace}\ q\ {\isasymLongrightarrow}\ p\ {\isasymrightarrow}\isactrlsub {\isasymsharp}{\isasymlparr}a\ {\isasymtriangleright}\ x{\isasymrparr}\ q%
\end{isabelle}

\item Communication:%
\begin{isabelle}%
p\ {\isasymrightarrow}\isactrlsub {\isasymflat}{\isasymlbrace}{\isasymtau}{\isasymrbrace}\ q\ {\isasymLongrightarrow}\ p\ {\isasymrightarrow}\isactrlsub {\isasymsharp}{\isasymlparr}{\isasymtau}{\isasymrparr}\ q%
\end{isabelle}%
\end{itemize}

\noindent For scope opening, we have a series of facts, one for each number of published channels. The
facts for one and two published channels are as follows:

\begin{itemize}%
\item One channel:%
\begin{isabelle}%
{\isasymlbrakk}p\ {\isasymrightarrow}\isactrlsub {\isasymflat}{\isasymlbrace}{\isasymnu}\ b{\isasymrbrace}\ Q\ b{\isacharsemicolon}\ {\isasymAnd}b{\isachardot}\ Q\ b\ {\isasymrightarrow}\isactrlsub {\isasymsharp}{\isasymlparr}a\ {\isasymtriangleleft}\ f\ b{\isasymrparr}\ R\ b{\isasymrbrakk}\ {\isasymLongrightarrow}\ p\ {\isasymrightarrow}\isactrlsub {\isasymsharp}{\isasymlparr}a\ {\isasymtriangleleft}\ {\isasymnu}\ b{\isachardot}\ f\ b{\isasymrparr}\ R\ b%
\end{isabelle}

\item Two channels:%
\begin{isabelle}%
{\isachardoublequote}{\isasymlbrakk}p\ {\isasymrightarrow}\isactrlsub {\isasymflat}{\isasymlbrace}{\isasymnu}\ b{\isasymrbrace}\ Q\ b{\isacharsemicolon}\ {\isasymAnd}b{\isachardot}\ Q\ b\ {\isasymrightarrow}\isactrlsub {\isasymsharp}{\isasymlparr}a\ {\isasymtriangleleft}\ {\isasymnu}\ c{\isachardot}\ f\ b\ c{\isasymrparr}\ R\ b\ c{\isasymrbrakk}\ {\isasymLongrightarrow}\\p\ {\isasymrightarrow}\isactrlsub {\isasymsharp}{\isasymlparr}a\ {\isasymtriangleleft}\ {\isasymnu}\ b\ c{\isachardot}\ f\ b\ c{\isasymrparr}\ R\ b\ c{\isachardoublequote}%
\end{isabelle}%
\end{itemize}

\noindent The facts for more published channels are analogous. All of these facts can be captured by a
single rule, which we do not show here for the sake of simplicity.

As it stands, the proper transition system has the issue that a scope can also be opened when the
respective channel is not published. For example, \isa{{\isasymnu}\ b{\isachardot}\ a\ {\isasymtriangleleft}\ x\ {\isasymrightarrow}\isactrlsub {\isasymsharp}{\isasymlparr}a\ {\isasymtriangleleft}\ {\isasymnu}\ b{\isachardot}\ x{\isasymrparr}\ {\isasymzero}} is a possible transition. We
are currently investigating ways to fix this issue. That said, this issue is of little relevance
for the rest of this paper, where we discuss the effects of transitions involving scope opening in
a way that is largely independent of the particularities of concrete transition systems.

A key issue with both the basic and the proper transition system is that, whenever a label
contains a binder, the scope of this binder includes any following target process. As a result, we
can treat neither of the two transition relations as a ternary relation, where source processes,
labels, and target processes are separate entities. As a solution, we consider the combination of
a label and an associated target process a single entity, which we call a \emph{residual}. Our
transition relations then become binary, relating source processes and residuals. This approach
has been taken in the formalization of \isa{{\isasympsi}}-calculi~\cite{bengtson:lmcs-7-1}, for example.

We define an inductive data type whose values are the residuals of the basic transition system.
There are two kinds of such residuals:

\begin{itemize}%
\item Acting:%
\begin{isabelle}%
{\isachardoublequote}Acting\ {\isasymalpha}\ p\ {\isasymequiv}\ {\isasymlbrace}{\isasymalpha}{\isasymrbrace}\ p{\isachardoublequote}%
\end{isabelle}

\item Opening:%
\begin{isabelle}%
{\isachardoublequote}Opening\ P\ {\isasymequiv}\ {\isasymlbrace}{\isasymnu}\ a{\isasymrbrace}\ P\ a{\isachardoublequote}%
\end{isabelle}%
\end{itemize}

\noindent Note that in the \isa{Opening} case we use HOAS and binder notation again.

Actually we do not just define a single data type for residuals but a type constructor
\isa{basic{\isacharunderscore}residual} that is parametrized  by the type of the target. As a result, terms
\isa{{\isasymlbrace}{\isasymxi}{\isasymrbrace}\ e} can be formed from terms \isa{e} of any type~\isa{{\isasymalpha}}, with the resulting type being
\isa{{\isasymalpha}\ basic{\isacharunderscore}residual}. This permits us to construct \emph{nested residuals}, residuals with two labels,
which have type \isa{process\ basic{\isacharunderscore}residual\ basic{\isacharunderscore}residual}. Nested residuals will play a role in
Subsect.~\ref{weak-residuals}.

We also introduce an analogous type constructor \isa{proper{\isacharunderscore}residual} for the proper transition
system. The definition of \isa{proper{\isacharunderscore}residual} is considerably more complex than the definition
of \isa{basic{\isacharunderscore}residual}, which is why we do not show it here. However, its general approach to
capturing scope opening is the same.%
\end{isamarkuptext}\isamarkuptrue%
\isadelimdocument
\endisadelimdocument
\isatagdocument
\isamarkupsubsection{Behavioral Equivalence%
}
\isamarkuptrue%
\endisatagdocument
{\isafolddocument}%
\isadelimdocument
\endisadelimdocument
\label{behavioral-equivalence}
\isadeliminvisible
\endisadeliminvisible
\isataginvisible
\isacommand{context}\isamarkupfalse%
\ transition{\isacharunderscore}system\ \isakeyword{begin}\isanewline
\isanewline
\isanewline
\isacommand{lemma}\isamarkupfalse%
\isanewline
\ \ \isakeyword{shows}\ {\isachardoublequoteopen}proper{\isachardot}sim\ {\isasymX}\ {\isasymLongrightarrow}\ {\isacharparenleft}{\isasymforall}p\ q\ {\isasymdelta}\ p{\isacharprime}{\isachardot}\ {\isasymX}\ p\ q\ {\isasymand}\ p\ {\isasymrightarrow}\isactrlsub {\isasymsharp}{\isasymlparr}{\isasymdelta}{\isasymrparr}\ p{\isacharprime}\ {\isasymlongrightarrow}\ {\isacharparenleft}{\isasymexists}q{\isacharprime}{\isachardot}\ q\ {\isasymrightarrow}\isactrlsub {\isasymsharp}{\isasymlparr}{\isasymdelta}{\isasymrparr}\ q{\isacharprime}\ {\isasymand}\ {\isasymX}\ p{\isacharprime}\ q{\isacharprime}{\isacharparenright}{\isacharparenright}{\isachardoublequoteclose}\isanewline
\ \ \isacommand{by}\isamarkupfalse%
\ {\isacharparenleft}blast\ elim{\isacharcolon}\ proper{\isacharunderscore}lift{\isacharunderscore}cases\ intro{\isacharcolon}\ simple{\isacharunderscore}lift{\isacharparenright}%
\endisataginvisible
{\isafoldinvisible}%
\isadeliminvisible
\endisadeliminvisible
\begin{isamarkuptext}%
Ultimately, we are interested in proving that different processes behave in the same way or at
least in similar ways. The standard notion of behavioral equivalence is \emph{bisimilarity}. A typical
approach to define bisimilarity is the following one:

\begin{enumerate}%
\item We define the predicate \isa{sim} on binary relations between processes as
follows:%
\begin{isabelle}%
sim\ {\isasymX}\ {\isasymlongleftrightarrow}\ {\isacharparenleft}{\isasymforall}p\ q\ {\isasymxi}\ p{\isacharprime}{\isachardot}\ {\isasymX}\ p\ q\ {\isasymand}\ p\ {\isasymrightarrow}{\isasymlparr}{\isasymxi}{\isasymrparr}\ p{\isacharprime}\ {\isasymlongrightarrow}\ {\isacharparenleft}{\isasymexists}q{\isacharprime}{\isachardot}\ q\ {\isasymrightarrow}{\isasymlparr}{\isasymxi}{\isasymrparr}\ q{\isacharprime}\ {\isasymand}\ {\isasymX}\ p{\isacharprime}\ q{\isacharprime}{\isacharparenright}{\isacharparenright}%
\end{isabelle}
A relation \isa{{\isasymX}} for which \isa{sim\ {\isasymX}} holds is called a \emph{simulation relation}.

\item We define the predicate \isa{bisim} on binary relations between processes as
follows:\footnote{Note that \isa{{\isacharunderscore}{\isasyminverse}{\isasyminverse}} is Isabelle/HOL syntax for conversion of relations that are
      represented by binary boolean functions.}%
\begin{isabelle}%
{\isachardoublequote}bisim\ {\isasymX}\ {\isasymlongleftrightarrow}\ sim\ {\isasymX}\ {\isasymand}\ sim\ {\isasymX}{\isasyminverse}{\isasyminverse}{\isachardoublequote}%
\end{isabelle}
A relation \isa{{\isasymX}} for which \isa{bisim\ {\isasymX}} holds is called a \emph{bisimulation relation}.

\item We define bisimilarity as the greatest bisimulation relation:%
\begin{isabelle}%
{\isacharparenleft}{\isasymsim}{\isacharparenright}\ {\isacharequal}\ {\isacharparenleft}GREATEST\ {\isasymX}{\isachardot}\ bisim\ {\isasymX}{\isacharparenright}%
\end{isabelle}%
\end{enumerate}

The above definition of \isa{sim} refers to labels and target processes separately and assumes
each transition has exactly one target process. This is a problem in the presence of scope
opening, where labels and target processes have to be considered together and where a single
transition may have different target processes depending on published channels.%
\end{isamarkuptext}\isamarkuptrue%
\isadeliminvisible
\endisadeliminvisible
\isataginvisible
\isacommand{end}\isamarkupfalse%
\endisataginvisible
{\isafoldinvisible}%
\isadeliminvisible
\endisadeliminvisible
\begin{isamarkuptext}%
Let us see how we can solve this problem for the basic transition system. We develop a definition
of the notion of simulation relation that retains the essence of the above definition but is able
to deal with the peculiarities of opening residuals. First, we define an operation
\isa{basic{\isacharunderscore}lift} that turns a relation between processes into a relation between basic
residuals. The general idea is that \isa{basic{\isacharunderscore}lift\ {\isasymX}} relates two residuals if and only if their
labels are the same and their target processes are in relation~\isa{{\isasymX}}. This idea can be tweaked
in an obvious way to work with opening residuals. We define \isa{basic{\isacharunderscore}lift} inductively using
the following rules:

\begin{itemize}%
\item Acting case:%
\begin{isabelle}%
{\isasymX}\ p\ q\ {\isasymLongrightarrow}\ basic{\isacharunderscore}lift\ {\isasymX}\ {\isacharparenleft}{\isasymlbrace}{\isasymalpha}{\isasymrbrace}\ p{\isacharparenright}\ {\isacharparenleft}{\isasymlbrace}{\isasymalpha}{\isasymrbrace}\ q{\isacharparenright}%
\end{isabelle}

\item Opening case:%
\begin{isabelle}%
{\isacharparenleft}{\isasymAnd}a{\isachardot}\ {\isasymX}\ {\isacharparenleft}P\ a{\isacharparenright}\ {\isacharparenleft}Q\ a{\isacharparenright}{\isacharparenright}\ {\isasymLongrightarrow}\ basic{\isacharunderscore}lift\ {\isasymX}\ {\isacharparenleft}{\isasymlbrace}{\isasymnu}\ a{\isasymrbrace}\ P\ a{\isacharparenright}\ {\isacharparenleft}{\isasymlbrace}{\isasymnu}\ a{\isasymrbrace}\ Q\ a{\isacharparenright}%
\end{isabelle}%
\end{itemize}

\noindent Using \isa{basic{\isacharunderscore}lift}, we define the notion of simulation relation for the basic transition
system as follows:%
\begin{isabelle}%
{\isachardoublequote}basic{\isachardot}sim\ {\isasymX}\ {\isasymlongleftrightarrow}\ {\isacharparenleft}{\isasymforall}p\ q\ c{\isachardot}\ {\isasymX}\ p\ q\ {\isasymand}\ p\ {\isasymrightarrow}\isactrlsub {\isasymflat}\ c\ {\isasymlongrightarrow}\ {\isacharparenleft}{\isasymexists}d{\isachardot}\ q\ {\isasymrightarrow}\isactrlsub {\isasymflat}\ d\ {\isasymand}\ basic{\isacharunderscore}lift\ {\isasymX}\ c\ d{\isacharparenright}{\isacharparenright}{\isachardoublequote}%
\end{isabelle}

For the proper transition system, we can define a lifting operation \isa{proper{\isacharunderscore}lift} in an
analogous way. Afterwards we can define the notion of simulation relation for the proper
transition system in exactly the same way as for the basic transition system, except that we have
to replace \isa{basic{\isacharunderscore}lift} by \isa{proper{\isacharunderscore}lift}.%
\end{isamarkuptext}\isamarkuptrue%
\isadelimdocument
\endisadelimdocument
\isatagdocument
\isamarkupsection{Residuals Axiomatically%
}
\isamarkuptrue%
\endisatagdocument
{\isafolddocument}%
\isadelimdocument
\endisadelimdocument
\label{residuals-axiomatically}
\begin{isamarkuptext}%
As it stands, we have to develop the theory of bisimilarity separately for the basic and the
  proper transition system. This means, we have to essentially duplicate definitions of concepts
  like simulation relation, bisimulation relation, and bisimilarity and also proofs of various
  properties of these concepts. The reason is that these two transition systems use different
  notions of residual and consequently different lifting operations.

  However, we can develop the theory of bisimilarity also generically. We describe axiomatically
  what a lifting operation is and construct all definitions and proofs of our theory with reference
  to a lifting operation parameter that fulfills the respective axioms. Whenever we want our theory
  to support a new notion of residual, we just have to define a concrete lifting operation for it
  and prove that this lifting operation has the necessary properties.

  Note that this approach not only allows for a common treatment of the basic and the proper
  transition system but also captures transition systems of other process calculi. In particular, it
  also works with transition systems that do not allow scope opening, like CCS~\cite{milner:ccs},
  as there is a trivial lifting operation for such systems.%
\end{isamarkuptext}\isamarkuptrue%
\isadelimdocument
\endisadelimdocument
\isatagdocument
\isamarkupsubsection{Residuals in General%
}
\isamarkuptrue%
\endisatagdocument
{\isafolddocument}%
\isadelimdocument
\endisadelimdocument
\isadeliminvisible
\endisadeliminvisible
\isataginvisible
\isacommand{context}\isamarkupfalse%
\ residual\ \isakeyword{begin}%
\endisataginvisible
{\isafoldinvisible}%
\isadeliminvisible
\endisadeliminvisible
\begin{isamarkuptext}%
As indicated in Subsect.~\ref{behavioral-equivalence}, a lifting operation \isa{lift} should
generally behave such that \isa{lift\ {\isasymX}} relates two residuals if and only if their labels are the
same and their target processes are in relation~\isa{{\isasymX}}. The axioms for lifting operations should
be in line with this behavior and should at the same time be specific enough to allow us to
develop the theory of bisimilarity solely based on a lifting operation parameter. It turns out
that the following axioms fulfill these requirements:\footnote{Note that \isa{{\isacharunderscore}\ OO\ {\isacharunderscore}} is Isabelle/HOL syntax
  for composition of relations that are represented by binary boolean functions.}

\begin{itemize}%
\item Equality preservation:%
\begin{isabelle}%
lift\ {\isacharparenleft}{\isacharequal}{\isacharparenright}\ {\isacharequal}\ {\isacharparenleft}{\isacharequal}{\isacharparenright}%
\end{isabelle}

\item Composition preservation:%
\begin{isabelle}%
lift\ {\isacharparenleft}{\isasymX}\ OO\ {\isasymY}{\isacharparenright}\ {\isacharequal}\ lift\ {\isasymX}\ OO\ lift\ {\isasymY}%
\end{isabelle}

\item Conversion preservation:%
\begin{isabelle}%
lift\ {\isasymX}{\isasyminverse}{\isasyminverse}\ {\isacharequal}\ {\isacharparenleft}lift\ {\isasymX}{\isacharparenright}{\isasyminverse}{\isasyminverse}%
\end{isabelle}%
\end{itemize}

The presence of the equality preservation and composition preservation axioms means that lifting
operations are functors. However, they are not functors in the Haskell sense. Haskell's functors
are specifically endofunctors on the category of types and functions, but lifting operations are
endofunctors on the category of types and \emph{relations}.\footnote{The analogy to functors in the Haskell
  sense can be seen from the fact that replacing \isa{lift}, \isa{{\isacharparenleft}{\isacharequal}{\isacharparenright}}, and \isa{{\isacharparenleft}OO{\isacharparenright}} in the
  equality preservation and composition preservation axioms by Haskell's \isatt{fmap}, \isatt{id}, and \isatt{(.)}
  yields Haskell's functor axioms.}

With the additional conversion preservation axiom, the axioms for lifting operations are precisely
the axioms for \emph{relators}~\cite[Sect.~5.1]{bird:aop}. Therefore, we can say that a residual
structure is just an endorelator on the category of types and relations -- no problem here.
Luckily, Isabelle/HOL automatically generates relator-specific constructs for every data type,
namely the lifting operation and various facts about it, including the instances of the axioms. As
a result, instantiating our theory of bisimilarity to a new notion of residual is extremely
simple.%
\end{isamarkuptext}\isamarkuptrue%
\isadeliminvisible
\endisadeliminvisible
\isataginvisible
\isacommand{end}\isamarkupfalse%
\endisataginvisible
{\isafoldinvisible}%
\isadeliminvisible
\endisadeliminvisible
\isadelimdocument
\endisadelimdocument
\isatagdocument
\isamarkupsubsection{Weak Residuals%
}
\isamarkuptrue%
\endisatagdocument
{\isafolddocument}%
\isadelimdocument
\endisadelimdocument
\label{weak-residuals}
\isadeliminvisible
\endisadeliminvisible
\isataginvisible
\isacommand{no{\isacharunderscore}notation}\isamarkupfalse%
\ proper{\isacharunderscore}transition\ {\isacharparenleft}{\isachardoublequoteopen}{\isacharunderscore}\ {\isasymrightarrow}\isactrlsub {\isasymsharp}{\isacharunderscore}{\isachardoublequoteclose}\ {\isacharbrackleft}{\isadigit{5}}{\isadigit{1}}{\isacharcomma}\ {\isadigit{5}}{\isadigit{1}}{\isacharbrackright}\ {\isadigit{5}}{\isadigit{0}}{\isacharparenright}\isanewline
\isacommand{notation}\isamarkupfalse%
\ proper{\isacharunderscore}transition\ {\isacharparenleft}{\isachardoublequoteopen}{\isacharunderscore}\ {\isasymrightarrow}{\isacharunderscore}{\isachardoublequoteclose}\ {\isacharbrackleft}{\isadigit{5}}{\isadigit{1}}{\isacharcomma}\ {\isadigit{5}}{\isadigit{1}}{\isacharbrackright}\ {\isadigit{5}}{\isadigit{0}}{\isacharparenright}\isanewline
\isanewline
\isacommand{no{\isacharunderscore}notation}\isamarkupfalse%
\ proper{\isachardot}weak{\isacharunderscore}transition\ {\isacharparenleft}\isakeyword{infix}\ {\isachardoublequoteopen}{\isasymRightarrow}\isactrlsub {\isasymsharp}{\isachardoublequoteclose}\ {\isadigit{5}}{\isadigit{0}}{\isacharparenright}\isanewline
\isacommand{notation}\isamarkupfalse%
\ proper{\isachardot}weak{\isacharunderscore}transition\ {\isacharparenleft}{\isachardoublequoteopen}{\isacharunderscore}\ {\isasymRightarrow}{\isacharunderscore}{\isachardoublequoteclose}\ {\isacharbrackleft}{\isadigit{5}}{\isadigit{1}}{\isacharcomma}\ {\isadigit{5}}{\isadigit{1}}{\isacharbrackright}\ {\isadigit{5}}{\isadigit{0}}{\isacharparenright}\isanewline
\isanewline
\isacommand{abbreviation}\isamarkupfalse%
\isanewline
\ \ silent{\isacharunderscore}transition{\isacharunderscore}closure\ {\isacharcolon}{\isacharcolon}\ {\isachardoublequoteopen}process\ {\isasymRightarrow}\ process\ {\isasymRightarrow}\ bool{\isachardoublequoteclose}\isanewline
\ \ {\isacharparenleft}\isakeyword{infix}\ {\isachardoublequoteopen}{\isasymrightarrow}{\isasymlparr}{\isasymtau}{\isasymrparr}\isactrlsup {\isacharasterisk}\isactrlsup {\isacharasterisk}{\isachardoublequoteclose}\ {\isadigit{5}}{\isadigit{0}}{\isacharparenright}\isanewline
\isakeyword{where}\isanewline
\ \ {\isachardoublequoteopen}{\isacharparenleft}{\isasymrightarrow}{\isasymlparr}{\isasymtau}{\isasymrparr}\isactrlsup {\isacharasterisk}\isactrlsup {\isacharasterisk}{\isacharparenright}\ {\isasymequiv}\ {\isacharparenleft}{\isasymlambda}s\ t{\isachardot}\ s\ {\isasymrightarrow}{\isasymlparr}{\isasymtau}{\isasymrparr}\ t{\isacharparenright}\isactrlsup {\isacharasterisk}\isactrlsup {\isacharasterisk}{\isachardoublequoteclose}\isanewline
\isanewline
\isacommand{lemma}\isamarkupfalse%
\ silent{\isacharunderscore}weak{\isacharunderscore}transition{\isacharunderscore}def{\isacharcolon}\isanewline
\ \ \isakeyword{shows}\ {\isachardoublequoteopen}p\ {\isasymRightarrow}{\isasymlparr}{\isasymtau}{\isasymrparr}\ q\ {\isasymlongleftrightarrow}\ p\ {\isasymrightarrow}{\isasymlparr}{\isasymtau}{\isasymrparr}\isactrlsup {\isacharasterisk}\isactrlsup {\isacharasterisk}\ q{\isachardoublequoteclose}\isanewline
\isacommand{proof}\isamarkupfalse%
\isanewline
\ \ \isacommand{assume}\isamarkupfalse%
\ {\isachardoublequoteopen}p\ {\isasymRightarrow}{\isasymlparr}{\isasymtau}{\isasymrparr}\ q{\isachardoublequoteclose}\isanewline
\ \ \isacommand{then}\isamarkupfalse%
\ \isacommand{show}\isamarkupfalse%
\ {\isachardoublequoteopen}p\ {\isasymrightarrow}{\isasymlparr}{\isasymtau}{\isasymrparr}\isactrlsup {\isacharasterisk}\isactrlsup {\isacharasterisk}\ q{\isachardoublequoteclose}\isanewline
\ \ \isacommand{proof}\isamarkupfalse%
\ {\isacharparenleft}induction\ p\ {\isachardoublequoteopen}{\isasymlparr}{\isasymtau}{\isasymrparr}\ q{\isachardoublequoteclose}\ arbitrary{\isacharcolon}\ q{\isacharparenright}\isanewline
\ \ \ \ \isacommand{case}\isamarkupfalse%
\ strong{\isacharunderscore}transition\isanewline
\ \ \ \ \isacommand{then}\isamarkupfalse%
\ \isacommand{show}\isamarkupfalse%
\ {\isacharquery}case\isanewline
\ \ \ \ \ \ \isacommand{by}\isamarkupfalse%
\ {\isacharparenleft}fact\ r{\isacharunderscore}into{\isacharunderscore}rtranclp{\isacharparenright}\isanewline
\ \ \isacommand{next}\isamarkupfalse%
\isanewline
\ \ \ \ \isacommand{case}\isamarkupfalse%
\ silent{\isacharunderscore}transition\isanewline
\ \ \ \ \isacommand{then}\isamarkupfalse%
\ \isacommand{show}\isamarkupfalse%
\ {\isacharquery}case\isanewline
\ \ \ \ \ \ \isacommand{by}\isamarkupfalse%
\ {\isacharparenleft}blast\ elim{\isacharcolon}\ proper{\isacharunderscore}silent{\isachardot}cases{\isacharparenright}\isanewline
\ \ \isacommand{next}\isamarkupfalse%
\isanewline
\ \ \ \ \isacommand{case}\isamarkupfalse%
\ {\isacharparenleft}composed{\isacharunderscore}transition\ p\ {\isacharunderscore}\ q{\isacharparenright}\isanewline
\ \ \ \ \isacommand{then}\isamarkupfalse%
\ \isacommand{obtain}\isamarkupfalse%
\ u\ \isakeyword{where}\ {\isachardoublequoteopen}p\ {\isasymrightarrow}{\isasymlparr}{\isasymtau}{\isasymrparr}\isactrlsup {\isacharasterisk}\isactrlsup {\isacharasterisk}\ u{\isachardoublequoteclose}\ \isakeyword{and}\ {\isachardoublequoteopen}u\ {\isasymrightarrow}{\isasymlparr}{\isasymtau}{\isasymrparr}\isactrlsup {\isacharasterisk}\isactrlsup {\isacharasterisk}\ q{\isachardoublequoteclose}\isanewline
\ \ \ \ \ \ \isacommand{by}\isamarkupfalse%
\ {\isacharparenleft}blast\ elim{\isacharcolon}\ proper{\isacharunderscore}silent{\isachardot}cases\ proper{\isacharunderscore}lift{\isacharunderscore}cases{\isacharparenright}\isanewline
\ \ \ \ \isacommand{then}\isamarkupfalse%
\ \isacommand{show}\isamarkupfalse%
\ {\isacharquery}case\isanewline
\ \ \ \ \ \ \isacommand{by}\isamarkupfalse%
\ {\isacharparenleft}fact\ rtranclp{\isacharunderscore}trans{\isacharparenright}\isanewline
\ \ \isacommand{qed}\isamarkupfalse%
\isanewline
\isacommand{next}\isamarkupfalse%
\isanewline
\ \ \isacommand{assume}\isamarkupfalse%
\ {\isachardoublequoteopen}p\ {\isasymrightarrow}{\isasymlparr}{\isasymtau}{\isasymrparr}\isactrlsup {\isacharasterisk}\isactrlsup {\isacharasterisk}\ q{\isachardoublequoteclose}\isanewline
\ \ \isacommand{then}\isamarkupfalse%
\ \isacommand{show}\isamarkupfalse%
\ {\isachardoublequoteopen}p\ {\isasymRightarrow}{\isasymlparr}{\isasymtau}{\isasymrparr}\ q{\isachardoublequoteclose}\isanewline
\ \ \isacommand{proof}\isamarkupfalse%
\ induction\isanewline
\ \ \ \ \isacommand{case}\isamarkupfalse%
\ base\isanewline
\ \ \ \ \isacommand{then}\isamarkupfalse%
\ \isacommand{show}\isamarkupfalse%
\ {\isacharquery}case\isanewline
\ \ \ \ \ \ \isacommand{by}\isamarkupfalse%
\ {\isacharparenleft}blast\ intro{\isacharcolon}\ proper{\isacharunderscore}internal{\isacharunderscore}is{\isacharunderscore}silent\ proper{\isachardot}silent{\isacharunderscore}transition{\isacharparenright}\isanewline
\ \ \isacommand{next}\isamarkupfalse%
\isanewline
\ \ \ \ \isacommand{case}\isamarkupfalse%
\ step\isanewline
\ \ \ \ \isacommand{then}\isamarkupfalse%
\ \isacommand{show}\isamarkupfalse%
\ {\isacharquery}case\isanewline
\ \ \ \ \ \ \isacommand{by}\isamarkupfalse%
\ {\isacharparenleft}blast\ intro{\isacharcolon}\isanewline
\ \ \ \ \ \ \ \ proper{\isachardot}strong{\isacharunderscore}transition\isanewline
\ \ \ \ \ \ \ \ proper{\isacharunderscore}internal{\isacharunderscore}is{\isacharunderscore}silent\isanewline
\ \ \ \ \ \ \ \ proper{\isachardot}composed{\isacharunderscore}transition{\isacharparenright}\isanewline
\ \ \isacommand{qed}\isamarkupfalse%
\isanewline
\isacommand{qed}\isamarkupfalse%
\isanewline
\isanewline
\isacommand{lemma}\isamarkupfalse%
\ observable{\isacharunderscore}weak{\isacharunderscore}transition{\isacharunderscore}def{\isacharcolon}\isanewline
\ \ \isakeyword{fixes}\ {\isasymxi}\ {\isacharcolon}{\isacharcolon}\ proper{\isacharunderscore}action\isanewline
\ \ \isakeyword{assumes}\ {\isachardoublequoteopen}{\isasymxi}\ {\isasymnoteq}\ {\isasymtau}{\isachardoublequoteclose}\isanewline
\ \ \isakeyword{shows}\ {\isachardoublequoteopen}p\ {\isasymRightarrow}{\isasymlparr}{\isasymxi}{\isasymrparr}\ q\ {\isasymlongleftrightarrow}\ {\isacharparenleft}{\isasymexists}s\ t{\isachardot}\ p\ {\isasymRightarrow}{\isasymlparr}{\isasymtau}{\isasymrparr}\ s\ {\isasymand}\ s\ {\isasymrightarrow}{\isasymlparr}{\isasymxi}{\isasymrparr}\ t\ {\isasymand}\ t\ {\isasymRightarrow}{\isasymlparr}{\isasymtau}{\isasymrparr}\ q{\isacharparenright}{\isachardoublequoteclose}\isanewline
\isacommand{proof}\isamarkupfalse%
\isanewline
\ \ \isacommand{assume}\isamarkupfalse%
\ {\isachardoublequoteopen}p\ {\isasymRightarrow}{\isasymlparr}{\isasymxi}{\isasymrparr}\ q{\isachardoublequoteclose}\isanewline
\ \ \isacommand{then}\isamarkupfalse%
\ \isacommand{show}\isamarkupfalse%
\ {\isachardoublequoteopen}{\isasymexists}s\ t{\isachardot}\ p\ {\isasymRightarrow}{\isasymlparr}{\isasymtau}{\isasymrparr}\ s\ {\isasymand}\ s\ {\isasymrightarrow}{\isasymlparr}{\isasymxi}{\isasymrparr}\ t\ {\isasymand}\ t\ {\isasymRightarrow}{\isasymlparr}{\isasymtau}{\isasymrparr}\ q{\isachardoublequoteclose}\isanewline
\ \ \isacommand{proof}\isamarkupfalse%
\ {\isacharparenleft}induction\ p\ {\isachardoublequoteopen}{\isasymlparr}{\isasymxi}{\isasymrparr}\ q{\isachardoublequoteclose}\ arbitrary{\isacharcolon}\ q{\isacharparenright}\isanewline
\ \ \ \ \isacommand{case}\isamarkupfalse%
\ strong{\isacharunderscore}transition\isanewline
\ \ \ \ \isacommand{then}\isamarkupfalse%
\ \isacommand{show}\isamarkupfalse%
\ {\isacharquery}case\isanewline
\ \ \ \ \ \ \isacommand{by}\isamarkupfalse%
\ {\isacharparenleft}blast\ intro{\isacharcolon}\ proper{\isacharunderscore}internal{\isacharunderscore}is{\isacharunderscore}silent\ proper{\isachardot}silent{\isacharunderscore}transition{\isacharparenright}\isanewline
\ \ \isacommand{next}\isamarkupfalse%
\isanewline
\ \ \ \ \isacommand{case}\isamarkupfalse%
\ silent{\isacharunderscore}transition\isanewline
\ \ \ \ \isacommand{with}\isamarkupfalse%
\ {\isacartoucheopen}{\isasymxi}\ {\isasymnoteq}\ {\isasymtau}{\isacartoucheclose}\ \isacommand{show}\isamarkupfalse%
\ {\isacharquery}case\isanewline
\ \ \ \ \ \ \isacommand{by}\isamarkupfalse%
\ {\isacharparenleft}blast\ elim{\isacharcolon}\ proper{\isacharunderscore}silent{\isachardot}cases{\isacharparenright}\isanewline
\ \ \isacommand{next}\isamarkupfalse%
\isanewline
\ \ \ \ \isacommand{case}\isamarkupfalse%
\ {\isacharparenleft}composed{\isacharunderscore}transition\ p\ {\isacharunderscore}\ q{\isacharparenright}\isanewline
\ \ \ \ \isacommand{then}\isamarkupfalse%
\ \isacommand{consider}\isamarkupfalse%
\isanewline
\ \ \ \ \ \ u\ \isakeyword{and}\ s\ \isakeyword{and}\ t\ \isakeyword{where}\ {\isachardoublequoteopen}p\ {\isasymRightarrow}{\isasymlparr}{\isasymtau}{\isasymrparr}\ u{\isachardoublequoteclose}\ \isakeyword{and}\ {\isachardoublequoteopen}u\ {\isasymRightarrow}{\isasymlparr}{\isasymtau}{\isasymrparr}\ s{\isachardoublequoteclose}\ \isakeyword{and}\ {\isachardoublequoteopen}s\ {\isasymrightarrow}{\isasymlparr}{\isasymxi}{\isasymrparr}\ t{\isachardoublequoteclose}\ \isakeyword{and}\ {\isachardoublequoteopen}t\ {\isasymRightarrow}{\isasymlparr}{\isasymtau}{\isasymrparr}\ q{\isachardoublequoteclose}\ {\isacharbar}\isanewline
\ \ \ \ \ \ s\ \isakeyword{and}\ t\ \isakeyword{and}\ u\ \isakeyword{where}\ {\isachardoublequoteopen}p\ {\isasymRightarrow}{\isasymlparr}{\isasymtau}{\isasymrparr}\ s{\isachardoublequoteclose}\ \isakeyword{and}\ {\isachardoublequoteopen}s\ {\isasymrightarrow}{\isasymlparr}{\isasymxi}{\isasymrparr}\ t{\isachardoublequoteclose}\ \isakeyword{and}\ {\isachardoublequoteopen}t\ {\isasymRightarrow}{\isasymlparr}{\isasymtau}{\isasymrparr}\ u{\isachardoublequoteclose}\ \isakeyword{and}\ {\isachardoublequoteopen}u\ {\isasymRightarrow}\ {\isasymlparr}{\isasymtau}{\isasymrparr}\ q{\isachardoublequoteclose}\isanewline
\ \ \ \ \ \ \isacommand{by}\isamarkupfalse%
\ {\isacharparenleft}blast\ elim{\isacharcolon}\ proper{\isacharunderscore}silent{\isachardot}cases\ proper{\isacharunderscore}lift{\isacharunderscore}cases{\isacharparenright}\isanewline
\ \ \ \ \isacommand{then}\isamarkupfalse%
\ \isacommand{show}\isamarkupfalse%
\ {\isacharquery}case\isanewline
\ \ \ \ \ \ \isacommand{by}\isamarkupfalse%
\ cases\ {\isacharparenleft}blast\ intro{\isacharcolon}\ proper{\isacharunderscore}internal{\isacharunderscore}is{\isacharunderscore}silent\ proper{\isachardot}composed{\isacharunderscore}transition{\isacharparenright}{\isacharplus}\isanewline
\ \ \isacommand{qed}\isamarkupfalse%
\isanewline
\isacommand{next}\isamarkupfalse%
\isanewline
\ \ \isacommand{assume}\isamarkupfalse%
\ {\isachardoublequoteopen}{\isasymexists}s\ t{\isachardot}\ p\ {\isasymRightarrow}{\isasymlparr}{\isasymtau}{\isasymrparr}\ s\ {\isasymand}\ s\ {\isasymrightarrow}{\isasymlparr}{\isasymxi}{\isasymrparr}\ t\ {\isasymand}\ t\ {\isasymRightarrow}{\isasymlparr}{\isasymtau}{\isasymrparr}\ q{\isachardoublequoteclose}\isanewline
\ \ \isacommand{then}\isamarkupfalse%
\ \isacommand{show}\isamarkupfalse%
\ {\isachardoublequoteopen}p\ {\isasymRightarrow}{\isasymlparr}{\isasymxi}{\isasymrparr}\ q{\isachardoublequoteclose}\isanewline
\ \ \ \ \isacommand{by}\isamarkupfalse%
\ {\isacharparenleft}fastforce\ intro{\isacharcolon}\isanewline
\ \ \ \ \ \ proper{\isacharunderscore}internal{\isacharunderscore}is{\isacharunderscore}silent\isanewline
\ \ \ \ \ \ proper{\isachardot}strong{\isacharunderscore}transition\isanewline
\ \ \ \ \ \ proper{\isachardot}composed{\isacharunderscore}transition{\isacharparenright}\isanewline
\isacommand{qed}\isamarkupfalse%
\endisataginvisible
{\isafoldinvisible}%
\isadeliminvisible
\endisadeliminvisible
\begin{isamarkuptext}%
Our axiomatic treatment of lifting operations allows us to handle ordinary bisimilarity, which is
also known as \emph{strong bisimilarity}. In practice, however, we are more interested in \emph{weak
  bisimilarity}. Weak bisimilarity cares only about observable behavior; it treats internal
communication as silent and ignores it.

Normally, weak bisimilarity can be elegantly defined as the bisimilarity of the \emph{weak transition
  relation}~\isa{{\isacharparenleft}{\isasymRightarrow}{\isacharparenright}}, which is derived from the original transition relation~\isa{{\isacharparenleft}{\isasymrightarrow}{\isacharparenright}} using the
following equivalences:\footnote{The notation \isa{{\isacharunderscore}\ {\isasymrightarrow}{\isasymlparr}{\isasymtau}{\isasymrparr}\isactrlsup {\isacharasterisk}\isactrlsup {\isacharasterisk}\ {\isacharunderscore}} stands for the reflexive and transitive closure
  of \isa{{\isacharunderscore}\ {\isasymrightarrow}{\isasymlparr}{\isasymtau}{\isasymrparr}\ {\isacharunderscore}}.}

\begin{itemize}%
\item Silent:%
\begin{isabelle}%
{\isachardoublequote}p\ {\isasymRightarrow}{\isasymlparr}{\isasymtau}{\isasymrparr}\ q\ {\isasymlongleftrightarrow}\ p\ {\isasymrightarrow}{\isasymlparr}{\isasymtau}{\isasymrparr}\isactrlsup {\isacharasterisk}\isactrlsup {\isacharasterisk}\ q{\isachardoublequote}%
\end{isabelle}

\item Observable:%
\begin{isabelle}%
{\isachardoublequote}{\isasymxi}\ {\isasymnoteq}\ {\isasymtau}\ {\isasymLongrightarrow}\ p\ {\isasymRightarrow}{\isasymlparr}{\isasymxi}{\isasymrparr}\ q\ {\isasymlongleftrightarrow}\ {\isacharparenleft}{\isasymexists}s\ t{\isachardot}\ p\ {\isasymRightarrow}{\isasymlparr}{\isasymtau}{\isasymrparr}\ s\ {\isasymand}\ s\ {\isasymrightarrow}{\isasymlparr}{\isasymxi}{\isasymrparr}\ t\ {\isasymand}\ t\ {\isasymRightarrow}{\isasymlparr}{\isasymtau}{\isasymrparr}\ q{\isacharparenright}{\isachardoublequote}%
\end{isabelle}%
\end{itemize}

Unfortunately, the above definition of~\isa{{\isacharparenleft}{\isasymRightarrow}{\isacharparenright}} refers to a dedicated silent label and thus cannot
be applied to our setting, where we treat residuals as black boxes. To resolve this issue, we
modify the definition of~\isa{{\isacharparenleft}{\isasymRightarrow}{\isacharparenright}} such that it is based on two relations that together identify
silence. We define these relations differently for different notions of residual but specify their
general properties by a set of axioms.%
\end{isamarkuptext}\isamarkuptrue%
\isadeliminvisible
\endisadeliminvisible
\isataginvisible
\isacommand{definition}\isamarkupfalse%
\isanewline
\ \ basic{\isacharunderscore}fuse\ {\isacharcolon}{\isacharcolon}\ {\isachardoublequoteopen}{\isacharprime}p\ basic{\isacharunderscore}residual\ basic{\isacharunderscore}residual\ {\isasymRightarrow}\ {\isacharprime}p\ basic{\isacharunderscore}residual\ {\isasymRightarrow}\ bool{\isachardoublequoteclose}\isanewline
\isakeyword{where}\isanewline
\ \ {\isachardoublequoteopen}basic{\isacharunderscore}fuse\ {\isacharequal}\ basic{\isacharunderscore}silent{\isasyminverse}{\isasyminverse}\ {\isasymsqunion}\ basic{\isacharunderscore}lift\ basic{\isacharunderscore}silent{\isasyminverse}{\isasyminverse}{\isachardoublequoteclose}\isanewline
\isanewline
\isanewline
\isacommand{lemma}\isamarkupfalse%
\ basic{\isacharunderscore}silent{\isacharunderscore}converse{\isacharunderscore}naturality{\isacharcolon}\isanewline
\ \ \isakeyword{shows}\ {\isachardoublequoteopen}basic{\isacharunderscore}silent{\isasyminverse}{\isasyminverse}\ OO\ {\isasymX}\ {\isacharequal}\ basic{\isacharunderscore}lift\ {\isasymX}\ OO\ basic{\isacharunderscore}silent{\isasyminverse}{\isasyminverse}{\isachardoublequoteclose}\isanewline
\ \ \isacommand{by}\isamarkupfalse%
\ {\isacharparenleft}blast\isanewline
\ \ \ \ elim{\isacharcolon}\ basic{\isacharunderscore}silent{\isachardot}cases\ basic{\isacharunderscore}lift{\isacharunderscore}cases\isanewline
\ \ \ \ intro{\isacharcolon}\ basic{\isacharunderscore}internal{\isacharunderscore}is{\isacharunderscore}silent\ basic{\isacharunderscore}lift{\isacharunderscore}intros{\isacharparenright}\isanewline
\isanewline
\isacommand{lemma}\isamarkupfalse%
\ basic{\isacharunderscore}absorb{\isacharunderscore}from{\isacharunderscore}fuse{\isacharcolon}\isanewline
\ \ \isakeyword{shows}\ {\isachardoublequoteopen}basic{\isachardot}absorb\ {\isasymI}\ {\isacharequal}\ basic{\isacharunderscore}lift\ {\isasymI}\ OO\ basic{\isacharunderscore}fuse{\isachardoublequoteclose}\isanewline
\ \ \isacommand{unfolding}\isamarkupfalse%
\ basic{\isacharunderscore}fuse{\isacharunderscore}def\isanewline
\ \ \isacommand{by}\isamarkupfalse%
\ {\isacharparenleft}simp\ add{\isacharcolon}\ basic{\isacharunderscore}silent{\isacharunderscore}converse{\isacharunderscore}naturality\ basic{\isacharunderscore}residual{\isachardot}rel{\isacharunderscore}compp{\isacharparenright}%
\endisataginvisible
{\isafoldinvisible}%
\isadeliminvisible
\endisadeliminvisible
\begin{isamarkuptext}%
The first of the relations that identify silence relates each process with the residual that
extends this process with the silent label. For \isa{basic{\isacharunderscore}residual}, we define this relation
inductively using the following rule:%
\begin{isabelle}%
basic{\isacharunderscore}silent\ p\ {\isacharparenleft}{\isasymlbrace}{\isasymtau}{\isasymrbrace}\ p{\isacharparenright}%
\end{isabelle}
For \isa{proper{\isacharunderscore}residual} and other residual type constructors, we can define the corresponding
relation in an analogous way.

The second of the relations that identify silence relates each nested residual that contains the
silent label at least once with the ordinary residual that is obtained by dropping this label. For
\isa{basic{\isacharunderscore}residual}, we define this relation inductively using the following rules:

\begin{itemize}%
\item Silent--acting case:%
\begin{isabelle}%
{\isachardoublequote}basic{\isacharunderscore}fuse\ {\isacharparenleft}{\isasymlbrace}{\isasymtau}{\isasymrbrace}{\isasymlbrace}{\isasymalpha}{\isasymrbrace}\ p{\isacharparenright}\ {\isacharparenleft}{\isasymlbrace}{\isasymalpha}{\isasymrbrace}\ p{\isacharparenright}{\isachardoublequote}%
\end{isabelle}

\item Silent--opening case:%
\begin{isabelle}%
{\isachardoublequote}basic{\isacharunderscore}fuse\ {\isacharparenleft}{\isasymlbrace}{\isasymtau}{\isasymrbrace}{\isasymlbrace}{\isasymnu}\ a{\isasymrbrace}\ P\ a{\isacharparenright}\ {\isacharparenleft}{\isasymlbrace}{\isasymnu}\ a{\isasymrbrace}\ P\ a{\isacharparenright}{\isachardoublequote}%
\end{isabelle}

\item Acting--silent case:%
\begin{isabelle}%
{\isachardoublequote}basic{\isacharunderscore}fuse\ {\isacharparenleft}{\isasymlbrace}{\isasymalpha}{\isasymrbrace}{\isasymlbrace}{\isasymtau}{\isasymrbrace}\ p{\isacharparenright}\ {\isacharparenleft}{\isasymlbrace}{\isasymalpha}{\isasymrbrace}\ p{\isacharparenright}{\isachardoublequote}%
\end{isabelle}

\item Opening--silent case:%
\begin{isabelle}%
{\isachardoublequote}basic{\isacharunderscore}fuse\ {\isacharparenleft}{\isasymlbrace}{\isasymnu}\ a{\isasymrbrace}{\isasymlbrace}{\isasymtau}{\isasymrbrace}\ P\ a{\isacharparenright}\ {\isacharparenleft}{\isasymlbrace}{\isasymnu}\ a{\isasymrbrace}\ P\ a{\isacharparenright}{\isachardoublequote}%
\end{isabelle}%
\end{itemize}

\noindent For \isa{proper{\isacharunderscore}residual} and other residual type constructors, we can define the corresponding
relation in an analogous way.%
\end{isamarkuptext}\isamarkuptrue%
\isadeliminvisible
\endisadeliminvisible
\isataginvisible
\isacommand{no{\isacharunderscore}notation}\isamarkupfalse%
\ proper{\isacharunderscore}transition\ {\isacharparenleft}{\isachardoublequoteopen}{\isacharunderscore}\ {\isasymrightarrow}{\isacharunderscore}{\isachardoublequoteclose}\ {\isacharbrackleft}{\isadigit{5}}{\isadigit{1}}{\isacharcomma}\ {\isadigit{5}}{\isadigit{1}}{\isacharbrackright}\ {\isadigit{5}}{\isadigit{0}}{\isacharparenright}\isanewline
\isanewline
\isacommand{no{\isacharunderscore}notation}\isamarkupfalse%
\ proper{\isachardot}weak{\isacharunderscore}transition\ {\isacharparenleft}{\isachardoublequoteopen}{\isacharunderscore}\ {\isasymRightarrow}{\isacharunderscore}{\isachardoublequoteclose}\ {\isacharbrackleft}{\isadigit{5}}{\isadigit{1}}{\isacharcomma}\ {\isadigit{5}}{\isadigit{1}}{\isacharbrackright}\ {\isadigit{5}}{\isadigit{0}}{\isacharparenright}\isanewline
\isanewline
\isacommand{notation}\isamarkupfalse%
\ basic{\isacharunderscore}transition\ {\isacharparenleft}\isakeyword{infix}\ {\isachardoublequoteopen}{\isasymrightarrow}{\isachardoublequoteclose}\ {\isadigit{5}}{\isadigit{0}}{\isacharparenright}\isanewline
\isanewline
\isacommand{notation}\isamarkupfalse%
\ basic{\isachardot}weak{\isacharunderscore}transition\ {\isacharparenleft}\isakeyword{infix}\ {\isachardoublequoteopen}{\isasymRightarrow}{\isachardoublequoteclose}\ {\isadigit{5}}{\isadigit{0}}{\isacharparenright}\isanewline
\isanewline
\isacommand{notation}\isamarkupfalse%
\ basic{\isacharunderscore}lift\ {\isacharparenleft}{\isachardoublequoteopen}lift{\isachardoublequoteclose}{\isacharparenright}\isanewline
\isanewline
\isacommand{notation}\isamarkupfalse%
\ basic{\isacharunderscore}silent\ {\isacharparenleft}{\isachardoublequoteopen}silent{\isachardoublequoteclose}{\isacharparenright}\isanewline
\isanewline
\isacommand{notation}\isamarkupfalse%
\ basic{\isacharunderscore}fuse\ {\isacharparenleft}{\isachardoublequoteopen}fuse{\isachardoublequoteclose}{\isacharparenright}%
\endisataginvisible
{\isafoldinvisible}%
\isadeliminvisible
\endisadeliminvisible
\begin{isamarkuptext}%
We define the weak transition relation~\isa{{\isacharparenleft}{\isasymRightarrow}{\isacharparenright}} of a given transition relation~\isa{{\isacharparenleft}{\isasymrightarrow}{\isacharparenright}}
generically based on two parameters \isa{silent} and \isa{fuse}. The definition of~\isa{{\isacharparenleft}{\isasymRightarrow}{\isacharparenright}} is
inductive, using the following rules:

\begin{itemize}%
\item Strong transitions:%
\begin{isabelle}%
p\ {\isasymrightarrow}\ c\ {\isasymLongrightarrow}\ p\ {\isasymRightarrow}\ c%
\end{isabelle}

\item Empty transitions:%
\begin{isabelle}%
silent\ p\ c\ {\isasymLongrightarrow}\ p\ {\isasymRightarrow}\ c%
\end{isabelle}

\item Compound transitions:%
\begin{isabelle}%
{\isasymlbrakk}p\ {\isasymRightarrow}\ c{\isacharsemicolon}\ lift\ {\isacharparenleft}{\isasymRightarrow}{\isacharparenright}\ c\ z{\isacharsemicolon}\ fuse\ z\ d{\isasymrbrakk}\ {\isasymLongrightarrow}\ p\ {\isasymRightarrow}\ d%
\end{isabelle}%
\end{itemize}

As indicated above, the behavior of \isa{silent} and \isa{fuse} should generally be such that
\isa{silent} adds a silent label to a process and \isa{fuse} removes a silent label from a nested
residual. The following axioms are in line with this behavior and are at the same time specific
enough to allow us to develop the theory of weak bisimilarity solely based on the \isa{silent}
and \isa{fuse} parameters:

\begin{itemize}%
\item Silent naturality:%
\begin{isabelle}%
{\isasymX}\ OO\ silent\ {\isacharequal}\ silent\ OO\ lift\ {\isasymX}%
\end{isabelle}

\item Fuse naturality:%
\begin{isabelle}%
lift\ {\isacharparenleft}lift\ {\isasymX}{\isacharparenright}\ OO\ fuse\ {\isacharequal}\ fuse\ OO\ lift\ {\isasymX}%
\end{isabelle}

\item Left-neutrality:%
\begin{isabelle}%
silent\ OO\ fuse\ {\isacharequal}\ {\isacharparenleft}{\isacharequal}{\isacharparenright}%
\end{isabelle}

\item Right-neutrality:%
\begin{isabelle}%
lift\ silent\ OO\ fuse\ {\isacharequal}\ {\isacharparenleft}{\isacharequal}{\isacharparenright}%
\end{isabelle}

\item Associativity:%
\begin{isabelle}%
fuse\ OO\ fuse\ {\isacharequal}\ lift\ fuse\ OO\ fuse%
\end{isabelle}%
\end{itemize}

The above axioms are precisely the axioms for monads.\footnote{The analogy to monads in the Haskell sense
  can be seen from the fact that replacing \isa{lift},\isa{silent}, \isa{fuse}, \isa{{\isacharparenleft}{\isacharequal}{\isacharparenright}}, and
  \isa{{\isacharparenleft}OO{\isacharparenright}} in these axioms by Haskell's \isatt{fmap}, \isatt{return}, \isatt{join}, \isatt{id}, and \isatt{(.)} yields the
  naturality properties of \isatt{return} and \isatt{join}, which hold automatically because of
  parametricity~\cite{wadler:fpca-1989}, as well as Haskell's \isatt{join}-based monad axioms.}
Therefore, we can say that a weak residual structure is just a monad in the category of types and
relations -- a completely unproblematic specification.

The monadic approach to weak residuals is actually very general. In particular, it makes
non-standard notions of silence possible, for example, by allowing multiple silent labels. Despite
this generality, typical properties of weak bisimilarity can be proved generically. Concretely, we
have developed formal proofs of the following statements:

\begin{itemize}%
\item Weak bisimilarity is the same as ``mixed'' bisimilarity, a notion of bisimilarity where
ordinary transitions are simulated by weak transitions.

\item Strong bisimilarity is a subrelation of weak bisimilarity.%
\end{itemize}

Furthermore, the generic definition of the weak transition relation~\isa{{\isacharparenleft}{\isasymRightarrow}{\isacharparenright}} is simpler than
the traditional definition shown at the beginning of \hyperref[weak-residuals]{this subsection} in
that it does not distinguish between silent and observable transitions; this distinction is pushed
into the definitions of the \isa{silent} and \isa{fuse} relations of the individual notions of
weak residual. The simple structure of the definition of~\isa{{\isacharparenleft}{\isasymRightarrow}{\isacharparenright}} encourages a simple structure
of generic proofs about weak transitions.%
\end{isamarkuptext}\isamarkuptrue%
\isadelimdocument
\endisadelimdocument
\isatagdocument
\isamarkupsubsection{Normal Weak Residuals%
}
\isamarkuptrue%
\endisatagdocument
{\isafolddocument}%
\isadelimdocument
\endisadelimdocument
\begin{isamarkuptext}%
The monadic approach to weak residuals forces us to implement the two relations \isa{silent} and
\isa{fuse} and prove their properties for every notion of residual. This usually takes quite some
effort, in particular because the definition of the \isa{fuse} relation is typically non-trivial,
which also affects the proofs of its properties. The reward is that we can use non-standard
notions of silence. However, we rarely need this additional power, because we are usually fine
with having \emph{normal weak residuals}, weak residuals that use a dedicated label to indicate
silence. We introduce a more specific algebraic structure for normal weak residuals, which is much
easier to instantiate than the monad structure of arbitrary weak residuals.

We identify silence using just a \isa{silent} relation that has the following properties:

\begin{itemize}%
\item Naturality:%
\begin{isabelle}%
{\isasymX}\ OO\ silent\ {\isacharequal}\ silent\ OO\ lift\ {\isasymX}%
\end{isabelle}

\item Left-uniqueness and left-totality:%
\begin{isabelle}%
silent\ OO\ silent{\isasyminverse}{\isasyminverse}\ {\isacharequal}\ {\isacharparenleft}{\isacharequal}{\isacharparenright}%
\end{isabelle}

\item Right-uniqueness:%
\begin{isabelle}%
{\isachardoublequote}silent{\isasyminverse}{\isasyminverse}\ OO\ silent\ {\isasymle}\ {\isacharparenleft}{\isacharequal}{\isacharparenright}{\isachardoublequote}%
\end{isabelle}%
\end{itemize}

Note that in fact these axioms ensure that \isa{silent} identifies a single label, our silent
label. This shows that, although we do not have first-class labels explicitly, we can nevertheless
have first-class representations of those labels that do not involve scope opening.

From a \isa{silent} relation we can derive a relation \isa{fuse} as follows:%
\begin{isabelle}%
fuse\ {\isacharequal}\ silent{\isasyminverse}{\isasyminverse}\ {\isasymsqunion}\ lift\ silent{\isasyminverse}{\isasyminverse}%
\end{isabelle}
This derivation captures exactly the idea that \isa{fuse} removes a silent label from a nested
residual: since \isa{silent} adds a silent label, \isa{silent{\isasyminverse}{\isasyminverse}} removes a silent label, and
consequently \isa{lift\ silent{\isasyminverse}{\isasyminverse}} removes a silent label under another label.

A \isa{silent} relation with the above properties and the \isa{fuse} relation derived from it
together fulfill the monad axioms, which shows that normal weak residuals are in fact weak
residuals.%
\end{isamarkuptext}\isamarkuptrue%
\isadelimdocument
\endisadelimdocument
\isatagdocument
\isamarkupsection{Related Work%
}
\isamarkuptrue%
\endisatagdocument
{\isafolddocument}%
\isadelimdocument
\endisadelimdocument
\label{related-work}
\begin{isamarkuptext}%
We are not the first ones to formalize a process calculus using HOAS.
  Honsell~et~al.~\cite{honsell:tcs-253-2}, for example, define a HOAS-version of the
  $\pi$-calculus in Coq and prove considerable parts of its metatheory. Their formalization does not
  allow the construction of \emph{exotic terms}, that is, processes whose structure depends on data. In
  our formalization, we use exotic terms deliberately for branching. However, we actually want
  process structure to depend on ordinary data only; dependence on channels, especially local
  channels, is something we would like to prevent. The approach of Honsell~et~al. for ruling out
  exotic terms is to declare the type of channels as a parameter. Unfortunately, we cannot adopt
  this approach for our formalization, since the classical nature of Isabelle/HOL makes exotic terms
  possible even if the channel type is abstract.

  Röckl and Hirschkoff~\cite{roeckel:jfp-13-2} develop a HOAS-based implementation of the
  language of the $\pi$-calculus in Isabelle/HOL and show that it is adequate with respect to an
  ordinary, first-order implementation. They prove several syntactic properties but do not deal with
  transitions and bisimilarity at all. Their definition of processes includes exotic terms, but they
  define a separate wellformedness predicate that identifies those processes that are not exotic.

  Neither of the two works described uses an abstract theory of transition systems like we do.
  However, there is also no real demand for that, as these developments only deal with one or even
  no transition system.

  We use HOAS, because we can avoid the difficulties of name handling this way. Another approach is
  to keep names explicit but use nominal logic~\cite{pitts:ic-186-2} to make name handling
  easier. Bengtson follows this approach in his dissertation~\cite{bengtson:phd-thesis}. He
  formalizes several process calculi, namely CCS, the $\pi$-calculus, and $\psi$-calculi, in
  Isabelle/HOL, making use of its support for nominal logic.%
\end{isamarkuptext}\isamarkuptrue%
\isadelimdocument
\endisadelimdocument
\isatagdocument
\isamarkupsection{Summary and Outlook%
}
\isamarkuptrue%
\endisatagdocument
{\isafolddocument}%
\isadelimdocument
\endisadelimdocument
\label{summary-and-outlook}
\begin{isamarkuptext}%
We have presented the language and the operational semantics of the \isa{{\isasymnatural}}-calculus, a
  general-purpose process calculus embedded into functional host languages using HOAS. Since the
  operational semantics of the \isa{{\isasymnatural}}-calculus is defined using two transition systems, we have
  developed an abstract theory of transition systems to treat concepts like bisimilarity
  generically. We have formalized~\cite{jeltsch:ouroboros-formalization} large parts of the
  \isa{{\isasymnatural}}-calculus and our complete theory of transition systems in Isabelle/HOL.

  Because of our use of HOAS, the \isa{{\isasymnatural}}-calculus allows process structure to depend on channels. An
  important task for the future is the development of techniques that allow us to prevent
  channel-dependent behavior while continuing to use HOAS for expressing binding of names.

  We plan to very soon start using our process calculus for developing a formally verified
  implementation of the Ouroboros family of consensus protocols. Our hope is to gain valuable
  feedback about our process calculus work this way, which can potentially lead to improvements of
  the calculus and its implementation.%
\end{isamarkuptext}\isamarkuptrue%
\subsection*{Acknowledgements}
\begin{isamarkuptext}%
I want to thank my colleagues at Well-Typed and IOHK for their encouraging and helpful feedback on
  this work. Special thanks go to Javier D\'iaz for stress-testing the \isa{{\isasymnatural}}-calculus in his proofs
  of network equivalences, pointing me to some important related work, and proofreading this paper.
  Furthermore, I want to particularly thank Duncan Coutts and Philipp Kant for providing guidance
  and feedback concerning our efforts towards a verified implementation of the Ouroboros family of
  consensus protocols as well as Edsko de Vries for his help with process calculi and in particular
  for putting me on the HOAS path.%
\end{isamarkuptext}\isamarkuptrue%
\isadeliminvisible
\endisadeliminvisible
\isataginvisible
\isacommand{end}\isamarkupfalse%
\endisataginvisible
{\isafoldinvisible}%
\isadeliminvisible
\endisadeliminvisible
\end{isabellebody}%

\bibliographystyle{splncs04}
\bibliography{root}

\begin{thebibliography}{10}
\providecommand{\url}[1]{\texttt{#1}}
\providecommand{\urlprefix}{URL }
\providecommand{\doi}[1]{https://doi.org/#1}

\bibitem{badertscher:ccs-2018}
Badertscher, C., Ga\v{z}i, P., Kiayias, A., Russell, A., Zikas, V.: {Ouroboros
  Genesis}: Composable proof-of-stake blockchains with dynamic availability.
  In: Proceedings of the 2018 {ACM SIGSAC} Conference on Computer and
  Communications Security. pp. 913--930. ACM, New York (2018).
  \doi{10.1145/3243734.3243848},
  \url{https://iohk.io/en/research/library/papers/ouroboros-genesiscomposable-proof-of-stake-blockchains-with-dynamic-availability/}

\bibitem{bengtson:phd-thesis}
Bengtson, J.: Formalising Process Calculi. Ph.D. thesis, Uppsala universitet,
  Uppsala, Sweden (2010), \url{http://www.itu.dk/people/jebe/my-phd.html}

\bibitem{bengtson:lmcs-7-1}
Bengtson, J., Johansson, M., Parrow, J., Victor, B.: Psi-calculi: A framework
  for mobile processes with nominal data and logic. Logical Methods in Computer
  Science  \textbf{7}(1) (Mar 2011). \doi{10.2168/LMCS-7(1:11)2011},
  \url{https://arxiv.org/abs/1101.3262}

\bibitem{bird:aop}
Bird, R., de~Moor, O.: Algebra of Programming. Prentice Hall International
  Series in Computer Science, Prentice Hall, Upper Saddle River, New Jersey
  (Sep 1997)

\bibitem{boudol:inria-00076939}
Boudol, G.: Asynchrony and the pi-calculus. Tech. Rep. RR-1702, INRIA,
  Rocquencourt, France (May 1992), \url{https://hal.inria.fr/inria-00076939}

\bibitem{chakravarty:plutus}
Chakravarty, M., Kireev, R., MacKenzie, K., McHale, V., M\"uller, J., Nemish,
  A., Nester, C., Peyton~Jones, M., Thompson, S., Valentine, R., Wadler, P.:
  Functional blockchain contracts (May 2019),
  \url{https://iohk.io/en/research/library/papers/functional-blockchain-contracts/},
  unpublished draft

\bibitem{david:eurocrypt-2018}
David, B., Ga\v{z}i, P., Kiayias, A., Russell, A.: {Ouroboros Praos}: An
  adaptively-secure, semi-synchronous proof-of-stake blockchain. In:
  Buus~Nielsen, J., Rijmen, V. (eds.) Advances in Cryptology, Lecture Notes in
  Computer Science, vol. 10821, pp. 66--98. Springer, Berlin/Heidelberg,
  Germany (2018). \doi{10.1007/978-3-319-78375-8\_3},
  \url{https://iohk.io/en/research/library/papers/ouroboros-praosan-adaptively-securesemi-synchronous-proof-of-stake-protocol/}

\bibitem{honsell:tcs-253-2}
Honsell, F., Miculan, M., Scagnetto, I.: $\pi$-calculus in (co)inductive type
  theory. Theoretical Computer Science  \textbf{253}(2),  239--285 (Feb 2001).
  \doi{10.1016/S0304-3975(00)00095-5},
  \url{https://users.dimi.uniud.it/~marino.miculan/Papers/TCS99.pdf}

\bibitem{jeltsch:wflp-2019-source}
Jeltsch, W.: A process calculus for formally verifying blockchain consensus
  protocols (Nov 2019), \url{https://github.com/jeltsch/wflp-2019}, source code
  of this paper

\bibitem{jeltsch:ouroboros-formalization}
Jeltsch, W., D\'iaz, J.: Towards a formalization of the {Ouroboros} protocol
  family (Nov 2019),
  \url{https://github.com/input-output-hk/fm-ouroboros/tree/bbeec3136ae68e7bb6800680e216b12db6c1113a/Isabelle},
  current version of the source code of the formalization

\bibitem{kiayias:crypto-2017}
Kiayias, A., Russell, A., David, B., Oliynykov, R.: {Ouroboros}: A provably
  secure proof-of-stake blockchain protocol. In: Katz, J., Shacham, H. (eds.)
  Advances in Cryptology, Lecture Notes in Computer Science, vol. 10401, pp.
  357--388. Springer, Berlin/Heidelberg, Germany (2017).
  \doi{10.1007/978-3-319-63688-7\_12},
  \url{https://iohk.io/en/research/library/papers/ouroborosa-provably-secure-proof-of-stake-blockchain-protocol/}

\bibitem{lamela:isola-2018}
Lamela~Seijas, P., Thompson, S.: Marlowe: Financial contracts on blockchain.
  In: Margaria, T., Steffen, B. (eds.) Leveraging Applications of Formal
  Methods, Verification and Validation, Lecture Notes in Computer Science, vol.
  11247, pp. 356--375. Springer, Berlin/Heidelberg, Germany (2018).
  \doi{10.1007/978-3-030-03427-6\_27},
  \url{https://iohk.io/en/research/library/papers/marlowefinancial-contracts-on-blockchain/}

\bibitem{milner:ccs}
Milner, R.: Communication and Concurrency. Prentice Hall International Series
  in Computer Science, Prentice Hall, Upper Saddle River, New Jersey (Mar 1989)

\bibitem{milner:pi-calculus}
Milner, R.: Communicating and Mobile Systems: The $\pi$-Calculus. Cambridge
  University Press, Cambridge, England (May 1999)

\bibitem{pfenning:pldi-1988}
Pfenning, F., Elliott, C.: Higher-order abstract syntax. In: Proceedings of the
  {ACM SIGPLAN} 1988 Conference on Programming Language Design and
  Implementation. pp. 199--208. ACM, New York (1988). \doi{10.1145/53990.54010}

\bibitem{pitts:ic-186-2}
Pitts, A.M.: Nominal logic, a first order theory of names and binding.
  Information and Computation  \textbf{186}(2),  165--193 (Nov 2003).
  \doi{10.1016/S0890-5401(03)00138-X}

\bibitem{roeckel:jfp-13-2}
Röckl, C., Hirschkoff, D.: A fully adequate shallow embedding of the
  $\pi$-calculus in {Isabelle/HOL} with mechanized syntax analysis. Journal of
  Functional Programming  \textbf{13}(2),  415--451 (Mar 2003).
  \doi{10.1017/S0956796802004653}

\bibitem{wadler:fpca-1989}
Wadler, P.: Theorems for free! In: Proceedings of the Fourth International
  Conference on Functional Programming Languages and Computer Architecture. pp.
  347--359. ACM, New York (1989). \doi{10.1145/99370.99404}

\end{thebibliography}

\end{document}